\documentclass[journal]{IEEEtran}
\ifCLASSINFOpdf
  \usepackage[pdftex]{graphicx}
   \graphicspath{{../figures/}{../jpeg/}}
\else
\fi

\usepackage[cmex10]{amsmath}
\usepackage{amssymb}
\usepackage{bm}

\usepackage{algpseudocode}
\usepackage{algorithm}
\usepackage{array}
\usepackage{url}
\usepackage[noadjust]{cite}

\newtheorem{proposition}{Proposition}
\newtheorem{remark}{Remark}

\usepackage{fancyhdr}
\usepackage{amsfonts}
\usepackage{xfrac}
\usepackage{color}
\usepackage{pgfplots}
\pgfplotsset{compat=newest}           
\pgfplotsset{compat=1.17}
\usepackage{enumerate}

\usepackage{cite}
\usepackage{mathrsfs}
\usepackage{subcaption}
\usepackage{stfloats}
\usepackage{makecell}
\usepackage{comment}
\usepackage{afterpage}
\usepackage{tikz}
\usepackage{mathdots}
\usepackage{yhmath}
\usepackage{cancel}
\usepackage{color}
\usepackage{siunitx}
\usepackage{array}
\usepackage{multirow}
\usepackage{textcomp}
\usepackage{gensymb}
\usepackage{tabularx}
\usepackage{booktabs}
\usepackage{graphicx}
\usepackage{xcolor}
\usepackage{mathtools}
\newtagform{normalsize}[\normalsize]{\normalsize(}{\normalsize)}

\allowdisplaybreaks
\usetikzlibrary{decorations.markings,decorations.pathmorphing}

\begin{document}
    \title{Performance Analysis of Pinching-Antenna Systems}
  \author{Dimitrios~Tyrovolas,~\IEEEmembership{Member,~IEEE,} 
      Sotiris A. Tegos,~\IEEEmembership{Senior Member,~IEEE,} \\ Panagiotis D. Diamantoulakis,~\IEEEmembership{Senior Member,~IEEE,} Sotiris Ioannidis, \\
      Christos K.~Liaskos,~\IEEEmembership{Member,~IEEE} and~George K.~Karagiannidis,~\IEEEmembership{Fellow,~IEEE}

\thanks{D. Tyrovolas is with the Aristotle University of Thessaloniki, 54124 Thessaloniki, Greece and with Dienekes SI IKE Heraklion, Crete, Greece (tyrovolas@auth.gr)}
\thanks{S. A. Tegos, P. D. Diamantoulakis, and G. K. Karagiannidis are with the Aristotle University of Thessaloniki, 54124 Thessaloniki, Greece (tegosoti@auth.gr, padiaman@auth.gr, geokarag@auth.gr).}
\thanks{S. Ioannidis is with the Dept. of Electrical and Computer Engineering, Technical University of Chania, Chania, Greece and with Dienekes SI IKE Heraklion, Crete, Greece (sotiris@ece.tuc.gr).}
\thanks{C. K. Liaskos is with the Computer Science Engineering Department, University of Ioannina, Ioannina, and Foundation for Research and Technology Hellas (FORTH), Greece (cliaskos@ics.forth.gr).} 
}
\maketitle

\begin{abstract}
The sixth generation of wireless networks envisions intelligent and adaptive environments capable of meeting the demands of emerging applications such as immersive extended reality, advanced healthcare, and the metaverse. However, this vision requires overcoming critical challenges, including the limitations of conventional wireless technologies in mitigating path loss and dynamically adapting to diverse user needs.  Among the proposed reconfigurable technologies, pinching antenna systems (PASs) offer a novel way to turn path loss into a programmable parameter by using dielectric waveguides to minimize propagation losses at high frequencies. In this paper, we develop a comprehensive analytical framework that derives closed-form expressions for the outage probability and average rate of PASs while incorporating both free-space path loss and waveguide attenuation under realistic conditions. In addition, we characterize the optimal placement of pinching antennas to maximize performance under waveguide losses. Numerical results show the significant impact of waveguide losses on system performance, especially for longer waveguides, emphasizing the importance of accurate loss modeling. Despite these challenges, PASs consistently outperform conventional systems in terms of reliability and data rate, underscoring their potential to enable high-performance programmable wireless environments.
\end{abstract}


\IEEEpeerreviewmaketitle

\usetagform{normalsize}

\section{Introduction}
As wireless networks evolve toward the sixth generation (6G), both industry and academia are focusing on designing networks that function as intelligent entities capable of meeting the demands of applications such as immersive extended reality, advanced healthcare, and the metaverse \cite{Holographic, healthcare, 6GMetaverse}. These diverse use cases not only require ultra-reliable and high-capacity communications, but also require the wireless environment to dynamically adapt to multiple users with different goals and requirements. Achieving this vision, however, requires advanced resource allocation mechanisms to exploit the large bandwidths of high-frequency bands while addressing critical challenges such as congestion that can significantly degrade signal quality \cite{6GNetwork}. In this direction, programmable wireless environments (PWEs) have emerged as a fundamental paradigm in which electromagnetic wave propagation is dynamically reconfigured to enable precise and personalized wireless services \cite{PWELiaskos}. In particular, PWEs rely on advanced reconfigurable components that can shape wireless propagation where, when, and how it is needed. Consequently, the ability to transform the inherently stochastic nature of wireless communications into a software-defined process underscores the importance of identifying and developing reconfigurable technologies that will drive the realization of intelligent and adaptive wireless networks.

\subsection{Motivation}
To realize the PWE vision, significant research efforts have focused on the development of advanced reconfigurable technologies capable of shaping the propagation environment. Among these, reconfigurable intelligent surfaces (RISs) have emerged as a leading solution, designed to dynamically manipulate incident electromagnetic (EM) waves \cite{SRE2020}. Specifically, by strategically placing RISs in the environment, RIS functionalities such as beam steering, absorption, and diffusion can be used to tailor wireless propagation to user requirements, enabling precise control of wireless transmissions \cite{liaskosmagazine,EEAlexand,TegosTVT,zeris}. In addition, researchers have advanced RIS architectures to expand their capabilities, including active RIS, which amplifies incoming signals \cite{Active}, and simultaneous transmit and reflect RIS, which enjoys a wider field of view \cite{STARRIS}, beyond-diagonal RIS for enhanced wave manipulation capabilities \cite{BDRIS}, light-emitting RIS for simultaneous EM wave manipulation and precise localization by optical signals \cite{LERIS}, and zero-energy RIS that harvests energy from EM waves for self-sustained operation \cite{zeris}. In parallel, alternative reconfigurable technologies, such as fluid and movable antennas, focus on dynamically adapting transmitter and receiver technologies to environmental changes \cite{Fluid,Movable}. Specifically, fluid antennas utilize reconfigurable materials, such as liquid metals, that can be dynamically shifted within the antenna structure to adapt their electromagnetic properties, while movable antennas physically reposition their entire structure to improve channel conditions \cite{Fluid,Movable}. To this end, these technologies provide appropriate directions for improving communications performance and lay a solid foundation for the PWE vision.

While researchers have proposed reconfigurable technologies that can serve as key components of PWEs, these efforts have focused on improving the effective channel gain and mitigating its stochastic nature \cite{SRE2020,Fluid,Movable}. However, to fully realize the vision of PWEs where every aspect of wireless propagation is programmable, it is critical to go beyond channel gain and address other key parameters that critically affect wireless transmissions. Among these, path loss stands out as a fundamental factor, particularly in high-frequency bands where it dominates signal degradation \cite{6GNetwork}. For example, RISs inherently suffer from double path loss, as the signal must traverse both the transmitter-to-RIS and RIS-to-receiver links, significantly increasing signal attenuation \cite{SRE2020}. However, while features such as beam steering are used to mitigate this problem by maximizing channel gain, these measures constrain RISs, which are designed to perform multiple functionalities beyond beam steering, but operate suboptimally in terms of path loss \cite{XRRF,ISACRIS}. Similarly, fluid and moving antennas cannot configure path loss because their positional adjustments are limited to a few wavelengths, resulting in minimal improvement. This observation, of course, raises a critical question: \textit{Can path loss itself be turned into a reconfigurable parameter?} Addressing this challenge would open a new degree of freedom in PWE design, consistent with its ultimate goal of programming all wireless-related parameters to optimize transmission quality.

A promising answer to the question of transforming path loss into a reconfigurable parameter lies in the innovative concept of \textit{pinching antennas}, a technology introduced by DOCOMO in 2022 that uses dielectric waveguides to guide EM waves at high frequencies \cite{book,DOCOMO}. Specifically, pinching antennas use these waveguides as radiating elements that are activated by applying small dielectric particles at any desired point along the waveguide, allowing for the dynamic creation of radiating sites \cite{DOCOMO,pinchingMAG,Ding2024TCOM,TegosPinching}. This unique design turns path loss into a programmable parameter by allowing flexible adjustment of antenna placement along the waveguide, thus changing the distance between radiating points and users. In addition, this capability allows pinching antennas to establish adjustable line-of-sight (LoS) links even in complex or obstructed environments without the need for additional hardware, making them a practical solution for dynamically shaping wireless propagation \cite{pinchingMAG,TegosPinching}. To this end, combining the waveguide efficiency of dielectric waveguides with the adaptability of pinching antennas introduces a transformative component poised to play a major role in next-generation PWEs.

Few recent studies have explored the potential of pinching antennas as an innovative reconfigurable technology for next-generation wireless networks, highlighting their unique capabilities \cite{Ding2024TCOM,TegosPinching,NOMAPinch,PASS}. Specifically, the authors in \cite{Ding2024TCOM} derived closed-form expressions for the average rate of a pinching antenna system (PAS) with a single pinching antenna operating over a lossless waveguide, while also showing that in multi-user MISO scenarios, pinching antennas can reconfigure channels to achieve ideal performance limits by dynamically adjusting antenna placement. In addition, the authors of \cite{TegosPinching} focused on the uplink optimization of PASs with lossless waveguides, addressing challenges such as antenna positioning and resource allocation. In particular, \cite{TegosPinching} proposed a convex reformulation of the antenna positioning problem and derived a closed-form solution for resource allocation, showing that pinching antennas allows a balanced trade-off between throughput and fairness. Finally, in \cite{NOMAPinch}, the authors investigated the integration of NOMA with PASs and showed that NOMA-assisted PASs outperform OMA-based PASs by effectively leveraging the simultaneous activation of multiple pinching antennas along a dielectric waveguide. 

\subsection{Contribution}
Despite the promising results of previous studies, a comprehensive evaluation of the performance improvements offered by PASs over conventional systems requires the derivation of closed-form expressions for critical performance metrics that account for both free-space path loss and waveguide attenuation. Furthermore, most existing works primarily focus on lossless waveguides or specific configurations, leaving a significant gap in understanding the impact of realistic waveguide losses on system performance. In this work, we bridge this gap by providing closed-form expressions to evaluate the performance improvements offered by PASs over conventional systems, while accounting for both free-space path loss and waveguide attenuation. Specifically:
\begin{itemize}
    \item We derive, for the first time in the literature, closed-form expressions for the outage probability of PASs with a waveguide and a pinching antenna for both the presence and absence of waveguide losses, allowing a rigorous evaluation of their reliability.
    \item We characterize the average rate of a PAS in the presence of waveguide losses, quantifying how waveguide attenuation and deployment area affect the achievable data rates.
    \item We provide a detailed analysis of the optimal placement of a pinching antenna to maximize communication performance under waveguide losses, providing practical insights for system design.
\end{itemize}
Through the provided numerical results, we show that waveguide losses have a significant impact on system performance, especially as the waveguide length increases, underscoring the importance of accurately modeling these losses in system design. Furthermore, even when practical waveguide losses exist, our results verify the consistent superiority of PASs over conventional systems in terms of both reliability and data rate performance, further highlighting their transformative potential to overcome the limitations of conventional systems.
\subsection{Structure}
The remainder of the paper is organized as follows. The system model is described in Section II. The performance analysis is given in Section III, and our results are presented in Section IV. Finally, Section V concludes the paper.

\section{System Model}

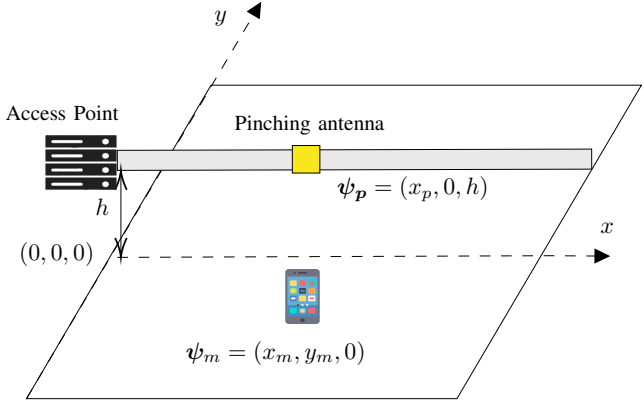
\begin{figure}
 \centering
    \resizebox{\columnwidth}{!}{
    \begin{tikzpicture}[x=0.75pt,y=0.75pt,yscale=-1,xscale=1]
        
        \draw   (161.22,50) -- (403.5,50) -- (299.66,226.6) -- (57.38,226.6) -- cycle ;
        \draw  [dash pattern={on 4.5pt off 4.5pt}]  (57.38,226.6) -- (187.97,5.58) ;
        \draw [shift={(189.5,3)}, rotate = 120.58] [fill={rgb, 255:red, 0; green, 0; blue, 0 }  ][line width=0.08]  [draw opacity=0] (8.93,-4.29) -- (0,0) -- (8.93,4.29) -- cycle    ;
        \draw  [fill={rgb, 255:red, 235; green, 232; blue, 233 }  ,fill opacity=1 ] (375.5,86) -- (108.47,86.7) -- (108.5,98) -- (375.53,97.3) -- cycle ;
        \draw  [fill={rgb, 255:red, 248; green, 231; blue, 28 }  ,fill opacity=1 ] (207,84) -- (222.5,84) -- (222.5,99.5) -- (207,99.5) -- cycle ;
        \draw (212.75,169) node  {\includegraphics[width=28.13pt,height=22.5pt]{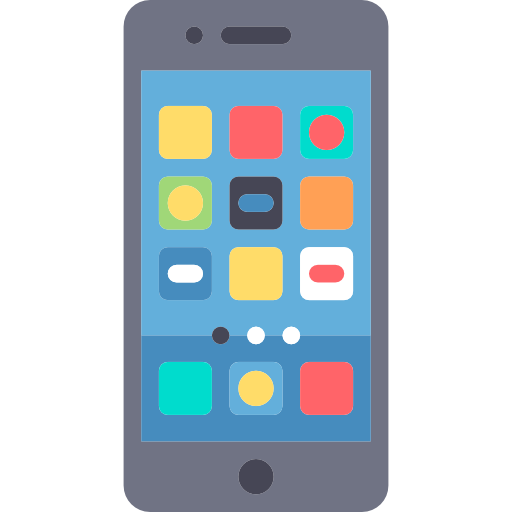}};
        \draw  [dash pattern={on 4.5pt off 4.5pt}]  (108.5,147) -- (351.78,146.07) -- (383,146.28) ;
        \draw [shift={(386,146.3)}, rotate = 180.39] [fill={rgb, 255:red, 0; green, 0; blue, 0 }  ][line width=0.08]  [draw opacity=0] (8.93,-4.29) -- (0,0) -- (8.93,4.29) -- cycle    ;
        \draw    (110.5,122) -- (110.5,100) ;
        \draw [shift={(110.5,98)}, rotate = 90] [color={rgb, 255:red, 0; green, 0; blue, 0 }  ][line width=0.75]    (10.93,-3.29) .. controls (6.95,-1.4) and (3.31,-0.3) .. (0,0) .. controls (3.31,0.3) and (6.95,1.4) .. (10.93,3.29)   ;
        \draw    (110.5,122) -- (110.5,145) ;
        \draw [shift={(110.5,147)}, rotate = 270] [color={rgb, 255:red, 0; green, 0; blue, 0 }  ][line width=0.75]    (10.93,-3.29) .. controls (6.95,-1.4) and (3.31,-0.3) .. (0,0) .. controls (3.31,0.3) and (6.95,1.4) .. (10.93,3.29)   ;
        \draw  [draw opacity=0] (104.25,147) .. controls (104.25,144.65) and (106.15,142.75) .. (108.5,142.75) .. controls (110.85,142.75) and (112.75,144.65) .. (112.75,147) .. controls (112.75,149.35) and (110.85,151.25) .. (108.5,151.25) .. controls (106.15,151.25) and (104.25,149.35) .. (104.25,147) -- cycle ;
        \draw (88,93.5) node  {\includegraphics[width=68.25pt,height=51.75pt]{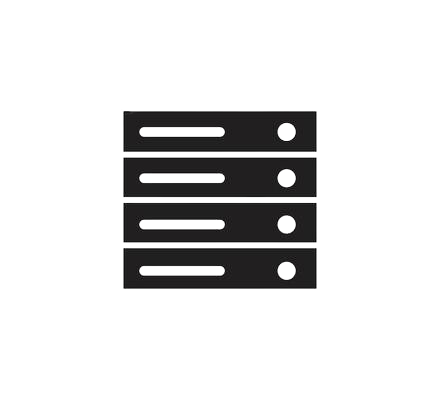}};
        
        \draw (52,136.99) node [anchor=north west][inner sep=0.75pt]    {$( 0,0,0)$};
        \draw (379.2,126.17) node [anchor=north west][inner sep=0.75pt]    {$x$};
        \draw (162,6.72) node [anchor=north west][inner sep=0.75pt]    {$y$};
        \draw (230.9,100.26) node [anchor=north west][inner sep=0.75pt]    {$\boldsymbol{\psi }_{\boldsymbol{p}}{} =\left( x_{p} ,0,h\right)$};
        \draw (95,112.37) node [anchor=north west][inner sep=0.75pt]    {$h$};
        \draw (173.4,65.89) node [anchor=north west][inner sep=0.75pt]   [align=left] {{\small Pinching antenna}};
        \draw (145.7,192.89) node [anchor=north west][inner sep=0.75pt]    {$\boldsymbol{\psi }_{m} =\left( x_{m} ,y_{m} ,0\right)$};
        \draw (44.5,60) node [anchor=north west][inner sep=0.75pt]   [align=left] {{\small Access Point}};
    \end{tikzpicture}
    }
    \caption{Overview of pinching antenna system.}
    \label{fig:system_model}
\end{figure}

We consider the downlink communication scenario shown in Fig. \ref{fig:system_model}, which targets high data rate communication between an access point (AP) and a single system user equipped with a single antenna. Specifically, it is assumed that the user is randomly placed within a rectangular area in the $x$-$y$ plane with side lengths equal to $D_x$ and $D_y$, respectively, and its position is represented by $\boldsymbol{\psi_m} = (x_m, y_m, 0)$, where $x_m$ is uniformly distributed over $[0, D_x]$ and $y_m$ is uniformly distributed over $\left[-\frac{D_y}{2}, \frac{D_y}{2}\right]$.  To provide robust communication, the AP uses pinching antennas based on a dielectric waveguide, which allows the position of the antenna to be dynamically adjusted along the waveguide to optimize link quality. Specifically, unlike traditional wireless systems, a PAS enables efficient transmission of high-frequency signals within the waveguide with minimal loss, while allowing the signal to radiate from any point along the waveguide using a controlled "pinching" mechanism. In addition, the waveguide is positioned parallel to the $x$-axis at a height $h$ and spans a length equal to $D_x$. Considering that the position of the pinching antenna is given as $\boldsymbol{\psi_p} = (x_p, 0, h)$ with $x_p \in [0, D_x]$, the channel $h_1$ between a pinching antenna and the user can be expressed as
\begin{equation}
    h_1 = \frac{\sqrt{\eta} e^{-j \frac{2\pi}{\lambda} \|\boldsymbol{\psi_m} -\boldsymbol{\psi_p}\|}}{\|\boldsymbol{\psi_m} -\boldsymbol{\psi_p}\|},
\end{equation}
with $\eta= \frac{\lambda^2}{16 \pi^2}$ representing the path loss at a reference distance of 1 m, $\lambda$ denoting the free-space wavelength of the signal, $j$ is the imaginary unit, and $\|\cdot\|$ expressing the Euclidean norm. However, as the signal travels along the waveguide, its phase changes due to the propagation characteristics of the dielectric material. Specifically, the interaction of the signal with the dielectric core and surrounding material within the waveguide reduces its phase velocity, characterized by the effective refractive index $n_{\mathrm{eff}}$, which in turn determines the guided wavelength as $\lambda_g = \frac{\lambda}{n_{\mathrm{eff}}}$. Consequently, the signal transmitted by the pinching antenna undergoes an additional phase shift, given by
\begin{equation}
    h_2 =  e^{-j \frac{2\pi}{\lambda_g} \|\boldsymbol{\psi_p} -\boldsymbol{\psi_0}\|},
\end{equation}
with $\boldsymbol{\psi_0}= (0, 0, h)$ denoting the position of the waveguide feeding point. Finally,  the waveguide is characterized by an absorption coefficient $\alpha\in [0, +\infty)$, which accounts for the intrinsic power attenuation of the signal as it traverses the waveguide. Consequently, assuming that the power attenuates exponentially, the received signal by the user can be expressed as
\begin{equation}
    y_r = \sqrt{P_t e^{-\alpha \|\boldsymbol{\psi_p} -\boldsymbol{\psi_0}\|}} h_1 h_2 s + w_n,
\end{equation}
where $s$ is the transmitted signal, assumed to satisfy $\mathbb{E}[|s|^2]=1$ with $\mathbb{E}[\cdot]$ denoting expectation and $w_n$ is additive white Gaussian noise with zero mean and variance $\sigma^2$. Therefore, the received SNR $\gamma_r$ can be written as
\begin{equation}\label{SNR1}
    \gamma_r= \frac{\eta P_t e^{-\alpha \|\boldsymbol{\psi_p} -\boldsymbol{\psi_0}\|} \left|e^{-j \left( \frac{2\pi}{\lambda} \|\boldsymbol{\psi_m} -\boldsymbol{\psi_p}\| + \frac{2\pi}{\lambda_g} \|\boldsymbol{\psi_p} -\boldsymbol{\psi_0}\|\right)}\right|^2}{{\sigma}^2 \|\boldsymbol{\psi_m} -\boldsymbol{\psi_p}\|^2}.
\end{equation}
Considering that $\left|e^{-jx}\right|=1$, \eqref{SNR1} can be rewritten as
\begin{equation}\label{SNR2}
    \gamma_r= \frac{\eta P_t e^{-\alpha \|\boldsymbol{\psi_p} -\boldsymbol{\psi_0}\|}}{{\sigma}^2 \|\boldsymbol{\psi_m} -\boldsymbol{\psi_p}\|^2}.
\end{equation}

\section{Performance Analysis}

In this section, we derive expressions for key performance metrics, such as outage probability and average rate, to thoroughly assess the communication performance of a PAS when the pinching antenna is positioned at the location that minimizes its distance from the user, providing insights into the efficiency of this placement strategy. Then, we present analytical derivations for the optimal pinching antenna placement that maximizes the received SNR, providing insights into the trade-offs between straightforward placement and performance-driven optimization.

\subsection{Outage Probability}
While placing the pinching antenna at $x_m$ ensures minimal path loss, it does not guarantee uninterrupted communication. In this direction, the following proposition provides a closed-form expression for the outage probability of the considered PAS.
\begin{proposition}\label{Prop02}
The outage probability of a PAS with a single pinching antenna placed at $(x_m,0,h)$ and a single-antenna user whose location is described by coordinates $x_m$ and $y_m$ can be expressed as in Table \ref{TableOut}, with $C=\frac{\eta P_t}{\gamma_{\mathrm{thr}} \sigma^2}$ where $\gamma_{\mathrm{thr}}$ is the SNR threshold.
\begin{table*}[t]
\centering
\caption{Expressions and conditions for outage probability $P_{out}$\label{TableOut}}
\begin{tabular}{p{0.5\linewidth}||p{0.4\linewidth}}
\hline\hline
\textbf{Expression of $P_{out}$} & \textbf{Condition} \\\hline\hline
$1$ & $h^2 \geq C$ \\\hline
$1 + \frac{4}{\alpha D_x D_y} \left( h{\mathrm{tan}}^{-1}\left(\frac{\sqrt{C-h^2}}{h} \right) - \sqrt{C-h^2} \right)$ & $h^2 \leq C$, $h^2 \geq C - \frac{D_y^2}{4}$ and $h^2 \geq Ce^{-\alpha D_x}$ \\\hline
$1 + \frac{\mathrm{ln}\left( \frac{h^2}{C} + \frac{D_y^2}{4C} \right) - 2}{\alpha D_x} + \frac{4h{\mathrm{tan}}^{-1}\left(\frac{D_y}{2h} \right)}{\alpha D_x D_y}$ & $h^2 \leq C - \frac{D_y^2}{4}$ and $h^2 \geq Ce^{-\alpha D_x}$ \\\hline
$1+ \frac{4}{\alpha D_x D_y} \Bigg( \sqrt{Ce^{-\alpha D_x}-h^2} - h{\mathrm{tan}}^{-1}\left(\frac{\sqrt{Ce^{-\alpha D_x}-h^2}}{h} \right) -\sqrt{C-h^2}$ \newline $ + h{\mathrm{tan}}^{-1}\left(\frac{\sqrt{C-h^2}}{h} \right) \Bigg)$ & $h^2 \geq C - \frac{D_y^2}{4}$ and $h^2 \leq Ce^{-\alpha D_x}$ \\\hline
$1+\frac{4}{\alpha D_x D_y} \Bigg( \sqrt{Ce^{-\alpha D_x}-h^2} - h{\mathrm{tan}}^{-1}\left(\frac{\sqrt{Ce^{-\alpha D_x}-h^2}}{h} \right) - \frac{D_y}{2}$ \newline $ + h{\mathrm{tan}}^{-1}\left(\frac{D_y}{2h} \right) \Bigg) + \frac{\mathrm{ln}\left( \frac{h^2}{C} + \frac{D_y^2}{4C}\right)}{\alpha D_x}$ & $h^2 \leq C - \frac{D_y^2}{4}$, $h^2 \leq Ce^{-\alpha D_x}$ \text{and} $h^2 \geq Ce^{-\alpha D_x}-\frac{D_y^2}{4}$ \\\hline
$0$ & $h^2 \leq Ce^{-\alpha D_x}-\frac{D_y^2}{4}$ \\\hline\hline
\end{tabular}
\end{table*}
\end{proposition}
\begin{IEEEproof}
Considering that $x_p=x_m$ and that $P_{out}= \Pr(\gamma_r \leq \gamma_{\mathrm{thr}})$, the outage probability can be expressed as
    \begin{equation}
P_{out}= \Pr\left( \frac{\eta P_t e^{-\alpha x_m}}{\sigma^2 \left({y^2_m}+h^2\right)} \leq \gamma_{\mathrm{thr}}\right),
    \end{equation}
which can be rewritten as
    \begin{equation}\label{Pout01}
        P_{out}=\Pr \left( {y^2_m} \geq Ce^{-\alpha x_m}-h^2\right).
    \end{equation}
To evaluate $P_{\text{out}}$, we first identify the feasible region for $y_m$ based on the condition $y_m^2 \geq Ce^{-\alpha x_m} - h^2$, and then integrate over the joint probability density function of $x_m$ and $y_m$.
Therefore, from the aforementioned inequality we obtain the following cases:
\begin{itemize}
    \item If $x_m \geq -\frac{1}{\alpha} \mathrm{ln}\left( \frac{h^2}{C}\right)$, then $-\frac{D_y}{2} \leq y_m \leq \frac{D_y}{2}$,
    \item If $x_m \leq -\frac{1}{\alpha} \mathrm{ln}\left( \frac{h^2}{C}\right)$, then $-\frac{D_y}{2} \leq y_m \leq -\sqrt{Ce^{-\alpha x_m}-h^2}$ and $  \sqrt{Ce^{-\alpha x_m}-h^2}\leq y_m \leq\frac{D_y}{2}$.
\end{itemize}
Given that $x_m \in [0, D_x]$ and $y_m \in \left[ -\frac{D_y}{2}, \frac{D_y}{2} \right]$, and assuming the independence of $x_m$ and $y_m$, the joint probability density function of $x_m$ and $y_m$ is $f_{x_m, y_m}(x_m, y_m) = \frac{1}{D_x D_y}$. Using this distribution, the outage probability can be expressed as shown in \eqref{15} at the top of the page. Specifically, the integral $I_1$ is calculated as
\begin{figure*}[t]
\begin{equation}\label{15}
\small
\begin{split}
    &P_{out} = \mathbb{I}\left(h^2 
    \leq C\right)\mathbb{I}\left(h^2 
    \geq Ce^{-\alpha D_x}\right)\Bigg[\mathbb{I}\left(h^2 
    \leq C-\frac{D^2_y}{4}\right)\int_{\frac{-1}{\alpha} \mathrm{ln}\left( \frac{h^2+\frac{D^2_y}{4}}{C}\right)}^{\frac{-1}{\alpha} \mathrm{ln}\left( \frac{h^2}{C}\right)} 
    I_1\, dx_m + \mathbb{I}\left(h^2 
    \geq C-\frac{D^2_y}{4}\right)\int_{0}^{\frac{-1}{\alpha} \mathrm{ln}\left( \frac{h^2}{C}\right)} 
   I_1\, dx_m \\&+  \int_{\frac{-1}{\alpha} \mathrm{ln}\left( \frac{h^2}{C}\right)}^{D_x} \int_{-\frac{D_y}{2}}^{\frac{D_y}{2}} \frac{1}{D_x D_y}\, dy_m \, dx_m\Bigg]  + \mathbb{I}\left(h^2 
    \leq Ce^{-\alpha D_x}\right) \mathbb{I}\left(h^2 
    \geq Ce^{-\alpha D_x} -\frac{D^2_y}{4}\right) \Bigg[\mathbb{I}\left(h^2 
    \leq C-\frac{D^2_y}{4}\right) \int_{\frac{-1}{\alpha} \mathrm{ln}\left( \frac{h^2+\frac{D^2_y}{4}}{C}\right)}^{D_x} 
    I_1\, dx_m \\& +\mathbb{I}\left(h^2 
    \geq C-\frac{D^2_y}{4}\right) \int_{0}^{D_x} 
    I_1\, dx_m\Bigg] +\mathbb{I}\left(h^2 
    \geq C\right)
\end{split}
\end{equation}
\hrule
\end{figure*}
\begin{equation}
\small
I_1 = \int_{-\frac{D_y}{2}}^{-\sqrt{Ce^{-\alpha x_m} - h^2}} \frac{1}{D_x D_y} \, dy_m + \int_{\sqrt{Ce^{-\alpha x_m} - h^2}}^{\frac{D_y}{2}} \frac{1}{D_x D_y} \, dy_m,
\end{equation}
where $\mathbb{I}(\cdot)$ denotes the indicator function, defined as
\begin{equation}
    \mathbb{I}(X) =
    \begin{cases} 
        1, & \text{if } X \text{ is true}, \\ 
        0, & \text{otherwise}.
    \end{cases}
\end{equation}
From \eqref{15} it can be seen that to obtain $P_{out}$ we need to calculate two integrals, i) $I_i=\int_{w_1}^{w_2} I_1\, dx_m $ where $I_i$ appears in \eqref{15} multiple times with different integration limits, and ii) $I_{j}=\int_{\frac{-1}{\alpha} \mathrm{ln}\left( \frac{h^2}{C}\right)}^{D_x} \int_{-\frac{D_y}{2}}^{\frac{D_y}{2}} \frac{1}{D_x D_y}\, dy_m \, dx_m$. Therefore, following similar steps as in Appendix \ref{App:A}, we derive that
\begin{equation}\label{I_i}
\small
\begin{split}
    &I_i = \frac{w_2-w_1}{D_x} + \frac{4}{\alpha D_x D_y}\Bigg[ \sqrt{Ce^{-\alpha w_2}-h^2}-\sqrt{Ce^{-\alpha w_1}-h^2} \\& -h{\mathrm{tan}}^{-1}\left(\frac{\sqrt{Ce^{-\alpha w_2}-h^2}}{h}\right) + h{\mathrm{tan}}^{-1}\left(\frac{\sqrt{Ce^{-\alpha w_1}-h^2}}{h}\right)\Bigg],
\end{split}
\end{equation}
and
\begin{equation}\label{I_j}
    I_{j}= 1 + \frac{1}{\alpha D_x} \mathrm{ln}\left( \frac{h^2}{C}\right) ,
\end{equation}
and after some algebraic manipulations, $P_{out}$ is derived, completing the proof.
\end{IEEEproof}
\begin{remark}
    In the case where multiple pinching antennas are used for beamforming, the antennas are positioned near $\psi_p$ with an interdistance of the order of $\lambda/2$ to ensure constructive interference to the user. However, given their proximity and the small value of $\lambda$, the path losses for these antennas are nearly identical to that of the primary pinching antenna at $\psi_p$. Consequently, the outage probability of a PAS with multiple pinching antennas can be tightly approximated by the expression for the single pinching antenna case, with $C = \frac{\eta N P_t}{\gamma_{\mathrm{thr}} \sigma^2}$, where $N$ is the number of the pinching antennas.
\end{remark}

After deriving the expression for the outage probability, we now consider a special case where the absorption coefficient is $\alpha=0$. This assumption represents an idealized scenario in which the waveguide experiences no attenuation, allowing us to isolate the effects of the served area on the outage probability.
\begin{proposition}
The outage probability of a system consisting of a single pinching antenna on a lossless dielectric waveguide  at $(x_m,0,h)$ and a single antenna user whose location is described by the coordinates $x_m$ and $y_m$ can be expressed as 
\begin{equation}\label{pout_loss}
P_{out} =
\begin{cases} 
1, & \text{$h^2 \geq C$,} \\ 
1 - \frac{2}{D_y} \sqrt{C - h^2}, & \text{$h^2 \leq C \leq h^2+\frac{D_y^2}{4}$,} \\ 
0 & \text{$C \geq h^2+\frac{D_y^2}{4}$}.
\end{cases}
\end{equation}
\end{proposition}
\begin{IEEEproof}
    Considering that $x_p=x_m$ and \eqref{Pout01}, the outage probability for the case of a lossless waveguide can be written as
    \begin{equation}\label{Poutlossless}
        P_{out}=\Pr \left( {y^2_m} \geq C-h^2\right).
    \end{equation}
Similarly to Proposition \ref{Prop02}, we need to identify the feasible region for $y_m$ based on the condition $y_m^2 \geq C - h^2$, and then integrate over the probability density function of $y_m$. In this direction, considering that $y_m\in \left[-\frac{D_y}{2},\frac{D_y}{2} \right]$ and that $y^2_m\geq 0$, then $-\frac{D_y}{2} \leq y_m \leq -\sqrt{C-h^2}$ and $  \sqrt{C-h^2}\leq y_m \leq\frac{D_y}{2}$, thus the outage probability for the special case is given as
\begin{equation}
\begin{split}
P_{out} & = \mathbb{I}\left(h^2 \geq C\right)+\mathbb{I}\left(h^2 
    +\frac{D^2_y}{4} \geq C\right)\mathbb{I} \left(h^2 
    \leq C\right)  \\& \times \left(\int_{\sqrt{C-h^2}}^{\frac{D_y}{2}} \frac{1}{D_y} dy_m +\int_{-\frac{D_y}{2}}^{-\sqrt{C-h^2}} \frac{1}{D_y} dy_m \right).
\end{split}
\end{equation}
Thus, after some algebraic manipulations, we derive \eqref{pout_loss} which completes the proof.
\end{IEEEproof}
\subsection{Average rate}
To further quantify the data rate capabilities of the considered PAS, we examine the average rate, which reflects the average achievable rate of the system. In this direction, the following proposition provides a closed-form expression for the average rate, allowing a comprehensive evaluation of the potential of pinching antennas to provide high data rates.
\begin{proposition}
The average rate $R_p$ of a system with a single pinching antenna placed at $(x_m,0,h)$ and a single-antenna user whose location is described by coordinates $x_m$ and $y_m$ can be expressed as in \eqref{17}, where $A=\frac{\eta P_t}{\sigma^2}$, $\left[ f(x) \right]_{a}^b= f(b)-f(a)$, $\mathrm{L_{i2}}\left(\cdot \right)$ is the dilogarithm function, and $z\left(x,y\right)$ is given as in \eqref{z} at the top of the next page.
\begin{figure*}[ht!]
\begin{equation}\label{17}
    \begin{split}
     R_p &= \frac{1}{D_x D_y}\Bigg(\frac{D_y}{\alpha \mathrm{ln}2} \left(\mathrm{L_{i2}}\left(\frac{-Ae^{-\alpha D_x}}{h^2+\frac{D_y^2}{4}}\right) -\mathrm{L_{i2}}\left(\frac{-A}{h^2+\frac{D_y^2}{4}}\right)\right) -\frac{4hD_x}{\ln2}\tan^{-1}\!\left(\frac{D_y}{2h}\right)  - \frac{8}{\alpha \ln2} \Bigg[ \frac{D_y}{4}\ln\left(\frac{D^2_y}{4}+x^2\right)   \\ & \qquad +x\tan^{-1}\!\left(\frac{D_y}{2x}\right) +\frac{h}{2} \tan^{-1}\!\left(\frac{D_y}{2x} \right) \ln \left(\frac{x-h}{x+h} \right)+\frac{h}{4} \Big( z\left(x,h\right) - z\left(x,-h\right)\Big)\Bigg]_{\sqrt{A+h^2}}^{\sqrt{Ae^{-\alpha D_x}+h^2}} \Bigg)  .
    \end{split}
\end{equation}
\hrule
\end{figure*}
\begin{figure*}[ht!]
\begin{equation}\label{z}
    \begin{split}
     z(x,y) &= 2\ln\left(x-y\right) \left(\tan^{-1}\!\left(\frac{2x}{D_y}\right) - \tan^{-1}\!\left(\frac{2y}{D_y}\right)\right)    +2 \mathrm{Im}\left\{\mathrm{L_{i2}}\left(\frac{y-x}{y-j \frac{D_y}{2}}  \right) \right\}.
    \end{split}
\end{equation}
\hrule
\end{figure*}
\end{proposition}
\begin{IEEEproof}
Taking into account \eqref{SNR2}, the average rate of the considered PAS can be expressed as
\begin{equation}\label{c1}
    R_p= \mathbb{E}\left[ \log_2\left( 1+\frac{A e^{-\alpha x_m}}{y^2_m+h^2}\right)\right],
\end{equation}
with $\mathbb{E}[\cdot]$ denoting expectation. Since \eqref{c1} is an even function with respect to $y_m$, it can be written in integral form as follows
\begin{equation}
    R_p=\frac{2}{D_x D_y\ln2} \int_0^{D_x} \int_0^{\frac{D_y}{2}} \ln \left(1+ \frac{Ae^{-\alpha x_m}}{y^2_m+h^2} \right) dy_m \,dx_m,
\end{equation}
which, after some algebraic manipulations, can be rewritten as
\begin{equation}
    R_p= \frac{1}{D_x D_y} \left(\frac{D_y}{\ln2}I_A + \frac{4}{\ln2} I_B-\frac{4hD_x}{\ln2}\tan^{-1}\!\left(\frac{D_y}{2h}\right)\right),
\end{equation}
where $I_A$ and $I_B$ are equal to
\begin{equation}\label{IA0}
    I_A= \int_0^{D_x} \ln\left(1+ \frac{Ae^{-\alpha x_m}}{h^2+\frac{D^2_y}{4}}\right) \, dx_m ,
\end{equation}
and
\begin{equation}\label{IB0}
    I_B= \int_0^{D_x}\! \sqrt{Ae^{-\alpha x_m}+h^2} \tan^{-1}\!\left(\frac{D_y}{2\sqrt{Ae^{-\alpha x_m}+h^2}}\right)\, dx_m.
\end{equation}
Thus, by following similar steps as shown in Appendix \ref{App:B}, we obtain the following results
\begin{equation}\label{IAt}
    I_A= \frac{1}{\alpha} \left(\mathrm{L_{i2}}\left(-\frac{Ae^{-\alpha D_x}}{h^2+\frac{D_y^2}{4}}\right) -\mathrm{L_{i2}}\left(-\frac{A}{h^2+\frac{D_y^2}{4}}\right)\right),
\end{equation}
and
\begin{equation}\label{IBt}
\begin{split}
    &I_B= -\frac{2}{\alpha} \Bigg[ \frac{D_y\ln\left(\frac{D^2_y}{4}+x^2\right)   }{4}+\frac{h\tan^{-1}\!\left(\frac{D_y}{2x} \right) }{2} \ln \left(\frac{x-h}{x+h} \right) \\&+x\tan^{-1}\!\left(\frac{D_y}{2x}\right) +\frac{h\Big( z\left(x,h\right) - z\left(x,-h\right)\Big)}{4} \Bigg]_{\sqrt{A+h^2}}^{\sqrt{Ae^{-\alpha D_x}+h^2}},
\end{split}
\end{equation}
where, after some algebraic manipulations, \eqref{17} is derived, which concludes the proof.
\end{IEEEproof}
\begin{remark}
    Similar to Remark 1, the average rate for the case of multiple pinching antennas can be closely approximated by the mathematical expression for the single-antenna case with $A = \frac{\eta N P_t}{\sigma^2}$.
\end{remark}

Finally, the average rate of the considered system for the special case of a lossless waveguide is given in \cite[(6)]{Ding2024TCOM}.

\subsection{Optimal Position of Pinching Antenna}
Previously, we analyzed the outage probability and average rate considering that the pinching antenna is placed at the location where the distance between the antenna and the user is minimized. While this serves as an intuitive antenna placement strategy, it does not necessarily maximize $\gamma_r$, especially with waveguide losses, as attenuation affects signal propagation along the dielectric structure, potentially shifting the optimal pinching antenna position away from direct alignment with the user. As a result, identifying the precise antenna position that maximizes $\gamma_r$ is crucial for understanding the trade-offs between straightforward placement and performance-driven optimization. To this end, the following proposition provides a closed-form expression for the optimal placement of the pinching antenna, ensuring maximum $\gamma_r$ under the given network constraints.

\begin{proposition}
The received SNR $\gamma_r$ of a single-antenna user located at $\boldsymbol{\psi_m}$ in a PAS with a single pinching antenna is maximized when the $x$-coordinate of the pinching antenna is given by
\begin{equation}\label{optimal}
\small
x_p^* =
\begin{cases}
x_{o_1}, & \text{if } \Delta>0 \text{ and } x_{o_1}>0  \text{ and }  f(x_{o_1}) > f(0), \\[8pt]
x_{o_2}, & \text{if } \Delta=0  \text{ and } x_{o_2}>0, \\[8pt]
0, & \text{otherwise,}
\end{cases}
\end{equation}
where $\Delta=4-4\alpha^2\left(y^2_m+h^2\right)$, $x_{o_1} = x_m - \frac{1 - \sqrt{1 - \alpha^2 (y_m^2 + h^2)}}{\alpha}$, $x_{o_2} = x_m - \frac{1}{\alpha}$, and $f(x) = \frac{e^{-\alpha x}}{(x_m - x)^2 + y_m^2 + h^2}$.
\end{proposition}

\begin{IEEEproof}
According to \eqref{SNR2}, the optimal position $x_p^*$ of the pinching antenna is the one that maximizes $f(x_p)$, which is given as
\begin{equation}
    f(x_p) = \frac{e^{-\alpha \|\boldsymbol{\psi_p} -\boldsymbol{\psi_0}\|}}{\|\boldsymbol{\psi_m} -\boldsymbol{\psi_p}\|^2} =  \frac{e^{-\alpha x_p}}{(x_m - x_p)^2 + y_m^2 + h^2}.
\end{equation}
To determine the optimal position $x_p^*$, we need to calculate the critical points of $f(x_p)$ among which the maximum can be identified. Therefore, the derivative of $f(x_p)$ is equal to
\begin{equation}\label{derivative}
\small
    f'(x_p) = \frac{e^{-\alpha x_p}\left(-\alpha \left((x_m - x_p)^2 + y_m^2 + h^2 \right) + 2 (x_m - x_p)\right)}{\left( (x_m - x_p)^2 + y_m^2 + h^2\right)^2}.
\end{equation}
The roots of \eqref{derivative} are determined by solving the numerator, resulting in the quadratic equation
\begin{equation}\label{9}
    \alpha z^2 -2 z + \alpha(y_m^2 + h^2)=0,
\end{equation}
with $z=x_m - x_p$. Thus, to find the roots of \eqref{9}, we need to calculate the discriminant $\Delta$ of \eqref{9}. First, considering that $\Delta>0$, the roots of \eqref{9} can be expressed as
\begin{equation}\label{root1}
    x_{p_{1,2}}= x_m - \frac{1 \pm \sqrt{1-\alpha^2 \left(y_m^2 + h^2 \right)}}{\alpha}.
\end{equation}
To confirm which of these critical points gives a local maximum of $f(x_p)$, we need to check the sign of the second derivative $f''(x_p)$, since the denominator of \eqref{derivative} is strictly positive. In this direction, by observing \eqref{derivative}, we can see that $f'(x_p)= f(x_p) g(x_p)$, where $g(x_p)$ is equal to
\begin{equation}
    g(x_p) = \frac{-\alpha \left((x_m - x_p)^2 + y_m^2 + h^2 \right) + 2 (x_m - x_p)}{(x_m - x_p)^2 + y_m^2 + h^2}.
\end{equation}
Therefore, the second derivative is expressed as $f''(x_p) = f'(x_p) g(x_p) + f(x_p) g'(x_p)$, and since the critical points are the roots of $f'(x_p)$, the sign of $f''(x_p)$ at these points is determined by the term $f(x_p)g'(x_p)$, provided at the top of the page as \eqref{12}.
\begin{figure*}[t]
\begin{equation}\label{12}
    \begin{split}
     f(x_p)g'(x_p) &= \frac{e^{-\alpha x_p}}{(x_m - x_p)^2 + y_m^2 + h^2}  \\ &
     \times \frac{(2\alpha(x_m-x_p)-2)((x_m - x_p)^2 + y_m^2 + h^2)-2(x_m-x_p)\left(-\alpha \left((x_m - x_p)^2 + y_m^2 + h^2 \right) + 2 (x_m - x_p)\right)}{\left( (x_m - x_p)^2 + y_m^2 + h^2\right)^2}.
    \end{split}
\end{equation}
\hrule
\end{figure*}
Thus, by substituting $x_{p_{1}}$ and $x_{p_{2}}$ into \eqref{12}, we find that $x_{p_{2}}$ corresponds to the local maximum of \eqref{SNR2}, as it yields a negative value for \eqref{12}, while $x_{p_{1}}$ corresponds to the local minimum of \eqref{SNR2}, as it results in a positive value for \eqref{12}. 

After establishing that $x_{p_2}$ is a local maximum and setting $x_{o_1}=x_{p_2}$, the next step is to determine the global maximum $x^*_p$ over the feasible region. In this direction, it is necessary to compare the function values at $x_{o_1}$ and the feeding point $x_p = 0$ in order to identify the global maximum point. Specifically, for $x_{o_1}$ to be the global maximum, it must satisfy $f(x_{o_1}) > f(0)$ while also being within the physically feasible range $[0, D_x]$. If neither of these conditions is met, the optimal placement reduces to $x_p^* = 0$, ensuring that the pinching antenna remains on the waveguide. As a result, enforcing these constraints yields the first branch of \eqref{optimal}. Similarly, the second branch of \eqref{optimal} is derived for the case where $\Delta = 0$, where in this scenario \eqref{9} has a single root given by $x_{o_2} = x_m - \frac{1}{\alpha}$. Specifically, by analyzing the sign of the first derivative around $x_{o_2}$, we find that $f'(x_p)$ changes from positive to negative, confirming $x_{o_2}$ as a global maximum, provided that it lies within the deployment region $[0, D_x]$. Finally, if $\Delta < 0$, \eqref{9} has no real roots, meaning that $f(x_p)$ is strictly decreasing as $f'(x_p)$ remains negative for all $x_p$, making $x_p=0$ the optimal position of the pinching antenna, which concludes the proof.
\end{IEEEproof}

\begin{remark}
In practice, the absorption coefficient $\alpha$ for dielectric waveguides is very small, resulting in $\sqrt{1-\alpha^2 \left(y_m^2 + h^2 \right)} \approx 1$ and strictly positive $\Delta$ for typical indoor scenarios where $y_m$ is of the order of $10$ m and $h$ is less than 5 m \cite{book}. Consequently, the position $x_{o_1}$ can be reduced to $x_{o_1} = x_m$, and since $x_m \geq 0$, the optimal position $x_p^*$ simplifies to $x_p^* = x_m$, emphasizing the effectiveness of aligning the antenna directly with the user along the horizontal plane. However, even if placing the antenna at $x_m$ is a practical strategy, the general case derived in \eqref{optimal} remains essential to describe the impact of higher attenuation in PASs and to understand optimal placement under varying waveguide conditions.
\end{remark}

\section{Numerical Results}
In this section, we evaluate the performance of PASs and the accuracy and validity of the derived expressions. For consistency, the system parameters are chosen as in \cite{Ding2024TCOM}, where the noise power $\sigma^2$ is -90 dBm, the carrier frequency $f_c=28$ GHz, and the effective refractive index $n_{\mathrm{eff}}=1.4$. We also assume $\gamma_{\mathrm{thr}}=100$, $h=3$ m, and unless otherwise stated, the absorption coefficient is set as $\alpha=0.01$ and the dimensions of the area are set as $D_x \times D_y= 10 \times 10$ $\mathrm{m}^2$. Finally, to evaluate the accuracy of the derived expressions, we perform Monte Carlo simulations with $10^6$ realizations.

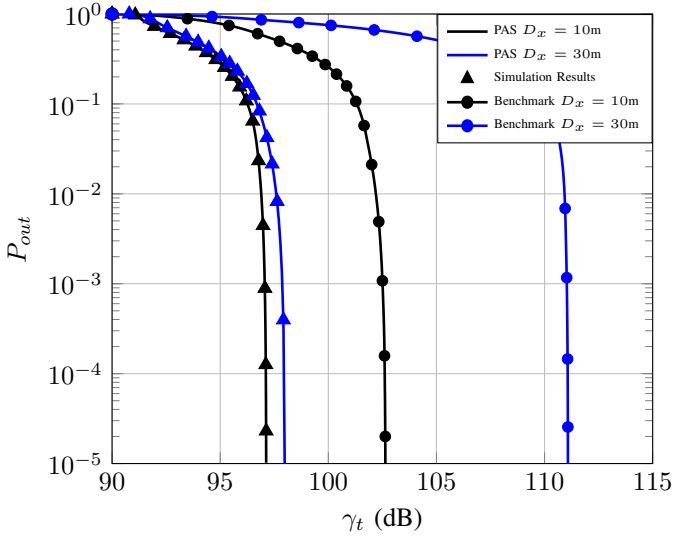
\begin{figure}
        \centering
        \begin{tikzpicture}
        \begin{semilogyaxis}[
            width=0.99\linewidth,
            xlabel = {$\gamma_t$ (dB)},
            ylabel = {$P_{out}$},
            xmin = 90, xmax = 115,
            ymin = 1e-5, ymax = 1,
            xtick = {90,95,...,115},
            grid = major,
            legend image post style={xscale=0.9, every mark/.append style={solid}},
            legend cell align = {left},
            legend style={
                at={(1,1)},
                anchor=north east,
                font = \tiny
            }
        ]
        
        \addplot[
            black,
            no marks,
            line width = 1pt,
            solid
        ]
        table {Outage_vs_pt/outpt_t_10.dat};
        \addlegendentry{PAS $D_x=10$m}
        
        \addplot[
            blue,
            no marks,
            line width = 1pt,
            solid
        ]
        table {Outage_vs_pt/outpt_t_30.dat};
        \addlegendentry{PAS $D_x=30$m}
                
        \addplot[
            black,
            only marks,
            mark=triangle*,
            mark size=3,
            mark repeat=1000,
        ]
        table {Outage_vs_pt/outpt_mc_10.dat};
        \addlegendentry{Simulation Results}

        \addplot[
            black,
            line width = 1pt,
            solid,
            mark=*,
            mark size = 2,
            mark repeat = 3000,
        ]
        table {Outage_vs_pt/outpt_b_10.dat};
        \addlegendentry{Benchmark $D_x=10$m}

        \addplot[
            blue,
            line width = 1pt,
            solid,
            mark=*,
            mark size = 2,
            mark repeat = 3000,
        ]
        table {Outage_vs_pt/outpt_b_30.dat};
        \addlegendentry{Benchmark $D_x=30$m}

        \addplot[
            black,
            only marks,
            mark=triangle*,
            mark size=3,
            mark indices={0,29,58,87,116,145,174,203,232,261,290,319,348,377,400, 410, 415, 417},
        ]
        table {Outage_vs_pt/outpt_mc_10.dat};

        \addplot[
            blue,
            only marks,
            mark=triangle*,
            mark size=3,
            mark indices={0, 3, 6, 9, 13, 16, 19, 23, 26, 29, 33, 36, 39, 43, 46, 49, 53, 56, 59, 63, 64, 65},
        ]
        table {Outage_vs_pt/outpt_mc_30.dat};

        \addplot[
            black,
            only marks,
            mark=*,
            mark size=2,
            mark indices={0,124,248,372,496,620,744,868,992,1116,1240,1364,1488,1612, 1680, 1720, 1735},
        ]
        table {Outage_vs_pt/outpt_b_10.dat};

        \addplot[
            blue,
            only marks,
            mark=*,
            mark size=2,
            mark indices={0, 20, 40, 64, 94, 154,248,372,496,620,744,868,992,1116,, 1117, 1236, 1257, 1269, 1273},
        ]
        table {Outage_vs_pt/outpt_b_30.dat};
        \end{semilogyaxis}
        \end{tikzpicture}
        \caption{Outage probability versus $\gamma_t$ for a PAS with one pinching antenna.}
        \label{Fig2}
\end{figure}

Fig. \ref{Fig2} illustrates the outage probability versus the transmit SNR, $\gamma_t = \frac{P_t}{\sigma^2}$, for a PAS with a single pinching antenna and a conventional system with a single antenna placed at the edge of the rectangular area at (0,0,0), for the cases where $D_x = 10$ m and $D_x = 30$ m. As can be seen, Fig. \ref{Fig2} clearly demonstrates the significant superiority of the PAS over the conventional setup in both deployment cases, as it consistently achieves lower outage probabilities for the same $\gamma_t$ values. This advantage is attributed to the flexibility of the pinching antennas to adjust their position along the dielectric waveguide to be closer to the user, effectively mitigating path loss and maintaining strong LoS communication links. In contrast, the fixed placement of the conventional antenna introduces significant large-scale path loss, particularly in larger areas, where the outage probability deteriorates significantly due to the increased average distance between the antenna and the user. For example, to achieve an outage probability of $10^{-5}$, the conventional system requires about 8 dB more transmit power for $D_x = 30$ m compared to $D_x = 10$ m, while the PAS shows negligible performance degradation between the two cases. Moreover, despite the presence of waveguide losses, the impact on the outage performance of the PAS is minimal because the dielectric waveguide ensures low propagation losses compared to free space propagation, underscoring the suitability of PASs for providing reliable communications. Finally, Fig. \ref{Fig2} highlights the close agreement between simulation results and theoretical predictions, thereby validating the accuracy of the derived analytical expressions and demonstrating their value for both system design and performance evaluation in practical deployments.

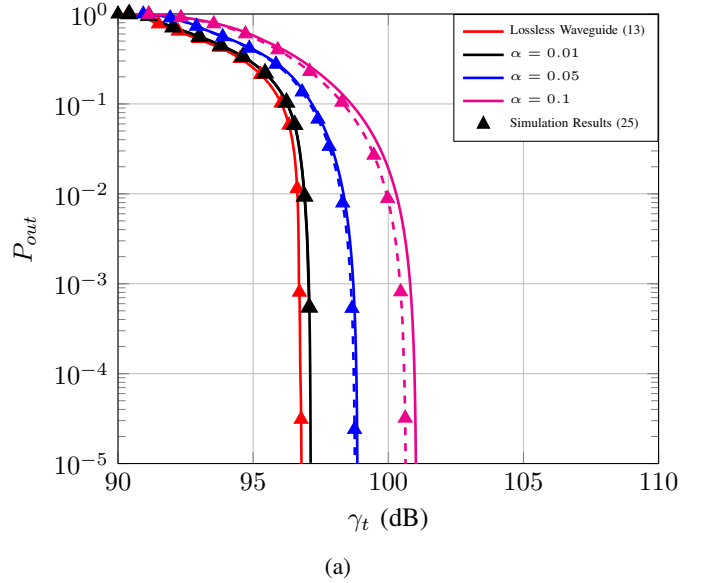
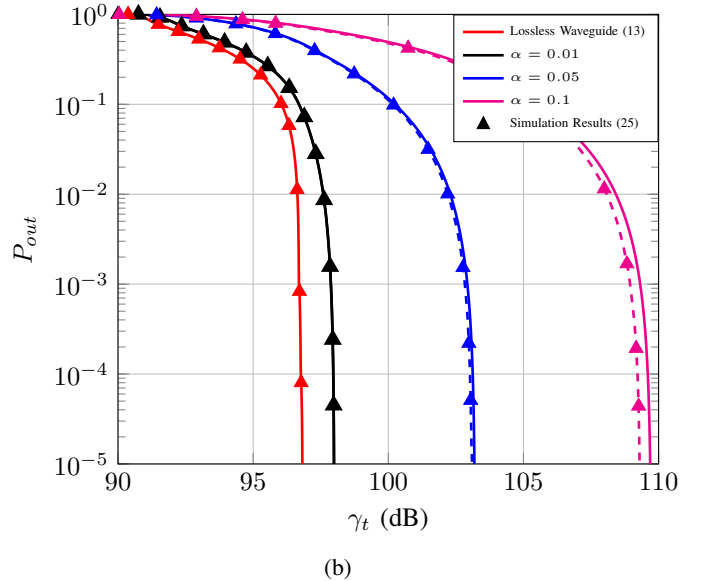
\begin{figure}[h!]
    \centering
    \begin{subfigure}{\linewidth}
        \centering
        \begin{tikzpicture}
        \begin{semilogyaxis}[
            width=0.99\linewidth,
            xlabel = {$\gamma_t$ (dB)},
            ylabel = {$P_{out}$},
            xmin = 90, xmax = 110,
            ymin = 1e-5, ymax = 1,
            xtick = {90,95,...,110},
            grid = major,
            legend image post style={xscale=0.9, every mark/.append style={solid}},
            legend cell align = {left},
            legend style={
                at={(1,1)},
                anchor=north east,
                font = \tiny
            }
        ]

        \addplot[
            red,
            no marks,
            line width = 1pt,
            solid
        ]
        table {Outage_beta/lossless_10.txt};
        \addlegendentry{Lossless Waveguide (13)}
        
        \addplot[
            black,
            no marks,
            line width = 1pt,
            solid
        ]
        table {Outage_beta/outbeta_t_10.dat};
        \addlegendentry{$\alpha=0.01$}
        
        \addplot[
            blue,
            no marks,
            line width = 1pt,
            solid
        ]
        table {Outage_beta/outbeta0.05_t_10.dat};
        \addlegendentry{$\alpha=0.05$}

        \addplot[
            magenta,
            no marks,
            line width = 1pt,
            solid
        ]
        table {Outage_beta/outbeta0.1_t_10.txt};
        \addlegendentry{$\alpha=0.1$}
                
        \addplot[
            black,
            only marks,
            mark=triangle*,
            mark size=3,
            mark repeat=1000,
        ]
        table {Outage_beta/outbeta0.01_mc_10.txt};
        \addlegendentry{Simulation Results (25)}
        
        \addplot[
            red,
            only marks,
            mark=triangle*,
            mark size = 3,
            mark indices={0, 10, 42, 68, 100, 138, 184, 238, 302, 330, 361, 370, 377},
        ]
        table {Outage_beta/lossless_mc_10.txt};

        \addplot[
            black,
            line width = 1pt,
            solid,
            mark=triangle*,
            mark size = 3,
            mark indices={0, 2, 4, 7, 11, 15, 20, 26, 33, 36, 40, 42},
            mark phase=1,
        ]
        table {Outage_beta/outbeta0.01_mc_10.txt};

        \addplot[
            blue,
            no marks,
            line width = 1pt,
            dashed,
        ]
        table {Outage_beta/outbeta0.05_mc_10.txt};
        
        \addplot[
            blue,
            only marks,
            mark=triangle*,
            mark size = 3,
            mark indices={0, 25, 57, 96, 145, 207, 285, 381, 450, 505, 579, 635, 651},
        ]
        table {Outage_beta/outbeta0.05_mc_10.txt};

        \addplot[
            magenta,
            no marks,
            line width = 1pt,
            dashed,
        ]
        table {Outage_beta/outbeta0.1_mc_10.txt};

        \addplot[
            magenta,
            only marks,
            mark=triangle*,
            mark size = 3,
            mark indices={0, 31, 72, 127, 197, 291, 413, 574,785, 895, 1012, 1056},
        ]
        table {Outage_beta/outbeta0.1_mc_10.txt};
        \end{semilogyaxis}
        \end{tikzpicture}
        \caption{}
    \end{subfigure}

    \vspace{0.5cm}

    \begin{subfigure}{\linewidth}
        \centering
        \begin{tikzpicture}
        \begin{semilogyaxis}[
            width=0.99\linewidth,
            xlabel = {$\gamma_t$ (dB)},
            ylabel = {$P_{out}$},
            xmin = 90, xmax = 110,
            ymin = 1e-5, ymax = 1,
            xtick = {90,95,...,110},
            grid = major,
            legend image post style={xscale=0.9, every mark/.append style={solid}},
            legend cell align = {left},
            legend style={
                at={(1,1)},
                anchor=north east,
                font = \tiny
            }
        ]

        \addplot[
            red,
            no marks,
            line width = 1pt,
            solid,
        ]
        table {Outage_beta/lossless_30.txt};
        \addlegendentry{Lossless Waveguide (13)}
        
        \addplot[
            black,
            no marks,
            line width = 1pt,
            solid
        ]
        table {Outage_beta/outbeta0.01_t_30.dat};
        \addlegendentry{$\alpha=0.01$}

        \addplot[
            blue,
            no marks,
            line width = 1pt,
            solid
        ]
        table {Outage_beta/outbeta0.05_t_30.dat};
        \addlegendentry{$\alpha=0.05$}

        \addplot[
            magenta,
            no marks,
            line width = 1pt,
            solid
        ]
        table {Outage_beta/outbeta0.1_t_30.dat};
        \addlegendentry{$\alpha=0.1$}                
        \addplot[
            black,
            only marks,
            mark=triangle*,
            mark size=3,
            mark repeat=1000,
        ]
        table {Outage_beta/outbeta0.01_mc_30.txt};
        \addlegendentry{Simulation Results (25)}

        \addplot[
            black,
            line width = 1pt,
            solid,
            mark=triangle*,
            mark size = 3,
            mark indices={0, 20, 44, 73, 108, 149, 199, 259, 331,390, 440, 480, 510, 524, 529},
        ]
        table {Outage_beta/outbeta0.01_mc_30.txt};

        \addplot[
            red,
            only marks,
            mark=triangle*,
            mark size = 3,
            mark indices={0, 10, 42, 68, 100, 138, 184, 238, 302, 330, 361, 370, 377},
        ]
        table {Outage_beta/lossless_mc_30.txt}; 

        \addplot[
            blue,
            no marks,
            line width = 1pt,
            dashed,
        ]
        table {Outage_beta/outbeta0.05_mc_30.txt};

        \addplot[
            blue,
            only marks,
            mark=triangle*,
            mark size = 3,
            mark indices={0,40,96,174,283,435,649,947,1305,1560, 1790, 1887, 1918},
        ]
        table {Outage_beta/outbeta0.05_mc_30.txt}; 

        \addplot[
            magenta,
            no marks,
            line width = 1pt,
            dashed
        ]
        table {Outage_beta/outbeta0.1_mc_30.txt}; 

        \addplot[
            magenta,
            only marks,
            mark=triangle*,
            mark size = 3,
            mark indices={1, 95, 189, 283, 514, 672, 879, 922, 942, 944},
        ]
        table{Outage_beta/outbeta0.1_mc_30.txt};
        
        \end{semilogyaxis}
        \end{tikzpicture}
        \caption{}
    \end{subfigure}
    \caption{Outage probability versus $\gamma_t$ for a PAS with one pinching antenna for various $\alpha$ values and a) $D_x=10$ m, and b) $D_x=30$ m.}
    \label{Fig3}
\end{figure}

Fig. \ref{Fig3}a and Fig. \ref{Fig3}b show the outage probability of a PAS with a single pinching antenna for various absorption coefficients $\alpha$, including the ideal case of a lossless waveguide. Fig. \ref{Fig3}a corresponds to the range of $D_x = 10 \, \mathrm{m}$, while Fig. \ref{Fig3}b represents the case of $D_x = 30 \, \mathrm{m}$. The $\alpha$ values studied range from $\alpha = 0.01$, representing low-loss materials, to $\alpha = 0.1$, describing dielectric waveguides with moderate losses \cite{book}. To further analyze the placement strategies, Fig. \ref{Fig3}a and Fig. \ref{Fig3}b present simulation results for the outage probability when the pinching antenna is optimally positioned, as derived in \eqref{optimal}, and compare these results with the theoretical expression given in Table \ref{TableOut}, which represents the case where the pinching antenna is placed at $(x_m,0,h)$ to minimize the path loss between the waveguide and the user. As can be observed, increasing $\alpha$ results in a degradation of the outage performance, especially as $D_x$ increases, due to the greater attenuation caused by the larger distance between the pinching antenna and the waveguide feedpoint. Nevertheless, the PAS consistently outperforms the conventional system shown in Fig. \ref{Fig2}, demonstrating its ability to provide reliable communication even when moderately lossy waveguides are employed. Furthermore, the divergence between the simulation results and the placement strategy that minimizes path loss becomes more pronounced as $\alpha$ increases, which is due to the dependence of the optimal placement on $\alpha$, as shown in \eqref{optimal}. Nevertheless, placing the pinching antenna at $(x_m, 0, h)$ remains a highly effective and practical design strategy that achieves near-optimal performance while simplifying the deployment of the pinching antenna. To this end, Figs. \ref{Fig3}a and \ref{Fig3}b highlight the critical importance of considering waveguide losses in the design of PASs, while also underscoring their unique ability to compensate for path loss even when typical losses within the waveguide are present. 

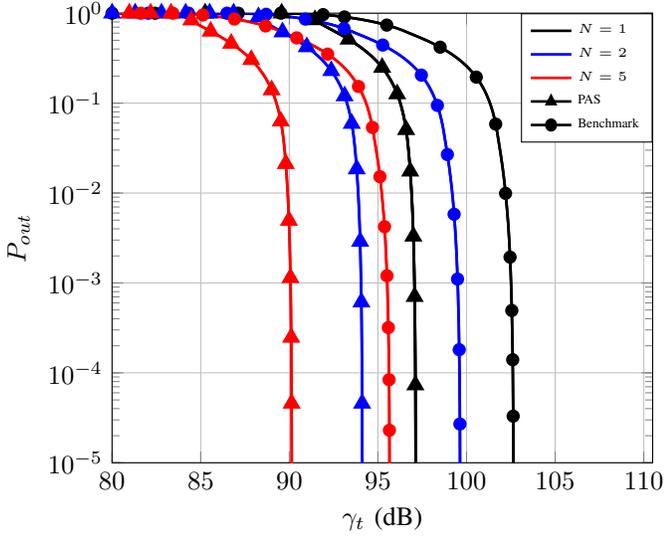
\begin{figure}
    \centering
    \begin{subfigure}{\linewidth}
        \centering
        \begin{tikzpicture}
        \begin{semilogyaxis}[
            width=0.99\linewidth,
            xlabel = {$\gamma_t$ (dB)},
            ylabel = {$P_{out}$},
            xmin = 80, xmax = 110.5,
            ymin = 1e-5, ymax = 1,
            xtick = {80,85,...,115},
            grid = major,
            legend image post style={xscale=0.9, every mark/.append style={solid}},
            legend cell align = {left},
            legend style={
                at={(1,1)},
                anchor=north east,
                font = \tiny
            }
        ]

        \addplot[
            black,
            no marks,
            line width = 1pt,
            solid,
        ]
        table {Outage_Multiple/out1.dat};
        \addlegendentry{$N=1$}

        \addplot[
            blue,
            no marks,
            line width = 1pt,
            solid,
        ]
        table {Outage_Multiple/out2.txt};
        \addlegendentry{$N=2$}

        \addplot[
            red,
            no marks,
            line width = 1pt,
            solid,
        ]
        table {Outage_Multiple/out5.dat};
        \addlegendentry{$N=5$}

        \addplot[
            black,
            line width = 1pt,
            solid,
            mark = triangle*,
            mark size = 2,
            mark repeat=5000,
        ]
        table {Outage_Multiple/out1.dat};
        \addlegendentry{PAS}

        \addplot[
            black,
            line width = 1pt,
            solid,
            mark=*,
            mark size = 2,
            mark repeat=4000,
        ]
        table {Outage_Multiple/outb1.dat};
        \addlegendentry{Benchmark}

        \addplot[
            blue,
            no marks,
            line width = 1pt,
            solid,
        ]
        table {Outage_Multiple/outb2.dat};

        \addplot[
            red,
            no marks,
            line width = 1pt,
            solid,
        ]
        table {Outage_Multiple/outb5.dat};

        \addplot[
            black,
            line width = 1pt,
            solid,
            mark=triangle*,
            mark size = 3,
            mark indices={45, 159, 257, 410, 646, 790, 890, 940, 980, 997, 1007, 1015, 1016, 1018},
            mark phase = 4,
        ]
        table {Outage_Multiple/out1.dat};

        \addplot[
            black,
            line width = 1pt,
            solid,
            mark=*,
            mark size = 2,
            mark indices= {0, 16,  84,  160, 291, 390, 690, 1400, 2250, 2900, 3300, 3500, 3580, 3623, 3645},
        ]
        table {Outage_Multiple/outb1.dat};

        \addplot[
            blue,
            line width = 1pt,
            solid,
            mark=triangle*,
            mark size = 3,
            mark indices= {0, 7, 18, 32, 51, 77, 114, 163, 231, 325, 390, 430, 460, 482, 490, 495},
            mark phase =2;
        ]
        table {Outage_Multiple/out2.dat};

        \addplot[
            blue,
            line width = 1pt,
            solid,
            mark=*,
            mark size = 2,
            mark indices= {0,13,35,70,129,227,388,654,1093, 1350, 1540, 1682, 1760, 1799, 1815}, 
        ]
        table {Outage_Multiple/outb2.dat};

        \addplot[
            red,
            line width = 1pt,
            solid,
            mark=triangle*,
            mark size = 3,
            mark indices= {0, 6, 14, 24, 37, 53, 75, 103, 140, 160, 172, 180, 184, 186, 187}, 
        ]
        table {Outage_Multiple/out5.dat};

        \addplot[
            red,
            line width = 1pt,
            solid,
            mark=*,
            mark size = 2,
            mark indices= {0, 10, 25, 46, 79, 128, 201, 310, 472, 570, 630, 670, 692, 705, 712, 715},
        ]
        table {Outage_Multiple/outb5.dat};

        \end{semilogyaxis}
        \end{tikzpicture}
    \end{subfigure}
    \caption{Outage probability versus $\gamma_t$ for a PAS with $N$ pinching antennas.}
    \label{Fig4}
\end{figure}

In Fig. \ref{Fig4}, the outage probability of a PAS with $N$ pinching antennas placed along a dielectric waveguide is shown and compared to a conventional system with an access point with $N$ antennas located at the waveguide feedpoint. As shown, the investigated PAS has a distinct advantage over the conventional system, as increasing $N$ significantly improves the outage performance due to the unique degree of freedom provided by the dynamic placement of the pinching antennas along the waveguide. In particular, the PAS with $N=1$ outperforms the conventional system with $N=2$, while the PAS with $N=2$ also outperforms the conventional system with $N=5$. As a result, Fig. \ref{Fig4} clearly demonstrates the superior efficiency of pinching antennas in utilizing the dielectric waveguide for communication and their ability to provide reliable communication with fewer antennas compared to conventional systems.

\begin{figure}
        \centering
        \begin{tikzpicture}
        \begin{axis}[
            width=0.99\linewidth,
            xlabel = {$\gamma_t$ (dB)},
            ylabel = {$R_p$},
            xmin = 77, xmax = 110,
            ymin = 0, ymax = 15,
            xtick = {70,80,...,110},
            ytick = {0,5,10,15},
            grid = major,
            legend image post style={xscale=0.9, every mark/.append style={solid}},
            legend cell align = {left},
            legend style={
                at={(0,1)},
                anchor=north west,
                font = \tiny
            }
        ]
        
        \addplot[
            black,
            no marks,
            line width = 1pt,
            solid
        ]
        table {Rate/reduced_rate10.dat};
        \addlegendentry{PAS $D_x=10$m}

        \addplot[
            blue,
            no marks,
            line width = 1pt,
            solid
        ]
        table {Rate/reduced_rate30.dat};
        \addlegendentry{PAS $D_x=30$m}

        \addplot[
            black,
            only marks,
            mark=triangle*,
            mark size=3,
            mark repeat = 2000,
        ]
        table {Rate/reduced_rate10.dat};
        \addlegendentry{Simulation Results}

        \addplot[
            black,
            line width = 1pt,
            solid,
            mark=*,
            mark size=2,
            mark repeat = 2000,
        ]
        table {Rate/reduced_rateb10.dat};
        \addlegendentry{Benchmark $D_x=10$m}
        
        \addplot[
            blue,
            line width = 1pt,
            solid,
            mark=*,
            mark size=2,
            mark repeat = 2000;
        ]
        table {Rate/reduced_rateb30.dat};
        \addlegendentry{Benchmark $D_x=30$m}

        \addplot[
            black,
            line width = 1pt,
            solid,
            mark=*,
            mark size=2,
            mark indices = {0, 1, 3, 8, 24, 73, 106, 129, 199},
        ]
        table {Rate/reduced_rateb10.dat};

        \addplot[
            black,
            line width = 1pt,
            solid,
            mark=triangle*,
            mark size=3,
            mark indices = {0, 2, 5, 15, 45, 99, 110, 150},
        ]
        table {Rate/reduced_rate10.dat};
        
        \addplot[
            blue,
            line width = 1pt,
            solid,
            mark=triangle*,
            mark size=3,
            mark indices = {0, 1, 3, 9, 25, 73, 106, 129, 199},
        ]
        table {Rate/reduced_rate30.dat};

        \addplot[
            blue,
            line width = 1pt,
            solid,
            mark=*,
            mark size=2,
            mark indices = {0, 1, 3, 8, 24, 74, 106, 129, 199},
        ]
        table {Rate/reduced_rateb30.dat};

        \end{axis}
        \end{tikzpicture}
        \caption{Average rate versus $\gamma_t$ for a PAS with one pinching antenna.}
        \label{Fig5}
\end{figure}
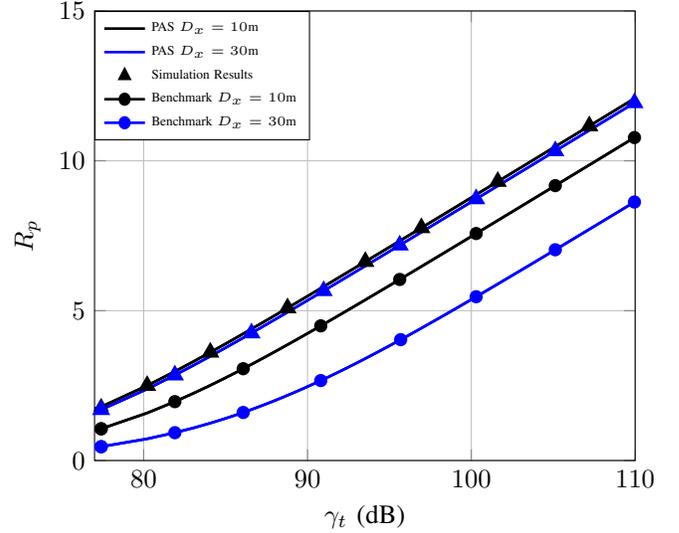

Fig. \ref{Fig5} illustrates the average rate as a function of $\gamma_t$ for a single-antenna PAS and a single-antenna conventional system, evaluated for deployment areas of $D_x = 10 \, \mathrm{m}$ and $D_x = 30 \, \mathrm{m}$. As observed, the PAS consistently achieves a higher average rate compared to the conventional system, demonstrating its ability to provide enhanced communication performance. In addition, the average rate of the PAS remains nearly identical for both deployment areas, with only minor differences due to waveguide losses, while the conventional system experiences a significant degradation when transitioning from $D_x = 10 \, \mathrm{m}$ to $D_x = 30 \, \mathrm{m}$. This behavior highlights the unique ability of pinching antennas to dynamically adjust their positions along the waveguide, compensating for path loss by minimizing the effective distance to the user.  In contrast, the conventional system suffers significant performance degradation due to the fixed placement of its antenna, which amplifies the path loss in wireless propagation as $D_x$ increases. Thus, Fig. \ref{Fig5} highlights the ability of pinching antennas to effectively harness the properties of dielectric waveguides to ensure consistent and robust performance in terms of data rate.

\begin{figure}
        \centering
        \begin{tikzpicture}
        \begin{axis}[
            width=0.99\linewidth,
            xlabel = {$\gamma_t$ (dB)},
            ylabel = {$R_p$},
            xmin = 77, xmax = 110,
            ymin = 0, ymax = 18,
            xtick = {70,80,...,110},
            ytick = {0,5,10,15,20},
            grid = major,
            legend image post style={xscale=0.9, every mark/.append style={solid}},
            legend cell align = {left},
            legend style={
                at={(0,1)},
                anchor=north west,
                font = \tiny
            }
        ]
        
        \addplot[
            black,
            no marks,
            line width = 1pt,
            solid
        ]
        table {Rate/reduced_rate10.dat};
        \addlegendentry{$N=1$}

        \addplot[
            blue,
            no marks,
            line width = 1pt,
            solid
        ]
        table {Rate_multi/reduced_rate2.dat};
        \addlegendentry{$N=2$}

        \addplot[
            red,
            no marks,
            line width = 1pt,
            solid
        ]
        table {Rate_multi/reduced_rate5.dat};
        \addlegendentry{$N=5$}

        \addplot[
            black,
            only marks,
            mark=triangle*,
            mark size=3,
            mark repeat = 1000,
        ]
        table {Rate/reduced_rate10.dat};
        \addlegendentry{PAS}

        \addplot[
            black,
            line width = 1pt,
            solid,
            mark=*,
            mark size=2,
            mark repeat = 1000,
        ]
        table {Rate_multi/reduced_rateb1.dat};
        \addlegendentry{Benchmark}
        
        \addplot[
            black,
            line width = 1pt,
            solid,
            mark=*,
            mark size=2,
            mark indices = {0, 0, 1, 3, 8, 20, 21, 25, 36, 69},
        ]
        table {Rate_multi/reduced_rateb1.dat};

        \addplot[
            black,
            line width = 1pt,
            solid,
            mark=triangle*,
            mark size=3,
            mark indices = {0, 1, 3, 6, 20, 56, 132, 180},
        ]
        table {Rate_multi/reduced_rate1.dat};
        
        \addplot[
            blue,
            line width = 1pt,
            solid,
            mark=triangle*,
            mark size=3,
            mark indices = {0, 1, 5, 12, 24, 38, 94, 240},
        ]
        table {Rate_multi/reduced_rate2.dat};

        \addplot[
            blue,
            line width = 1pt,
            solid,
            mark=*,
            mark size=2,
            mark indices = {0, 0, 1, 3, 8, 20, 21, 25, 36, 69},
        ]
        table {Rate_multi/reduced_rateb2.dat};
        
        \addplot[
            red,
            line width = 1pt,
            solid,
            mark=triangle*,
            mark size=3,
            mark indices = {0, 0, 1, 3, 8, 20, 21, 25, 36, 69},
        ]
        table {Rate_multi/reduced_rate5.dat};

        \addplot[
            red,
            line width = 1pt,
            solid,
            mark=*,
            mark size=2,
            mark indices = {0, 0, 1, 3, 8, 20, 21, 25, 36, 69},
        ]
        table {Rate_multi/reduced_rateb5.dat};
        
        \end{axis}
        \end{tikzpicture}
        \caption{Average rate versus $\gamma_t$ for a PAS with $N$ pinching antennas.}
        \label{Fig6}
\end{figure}
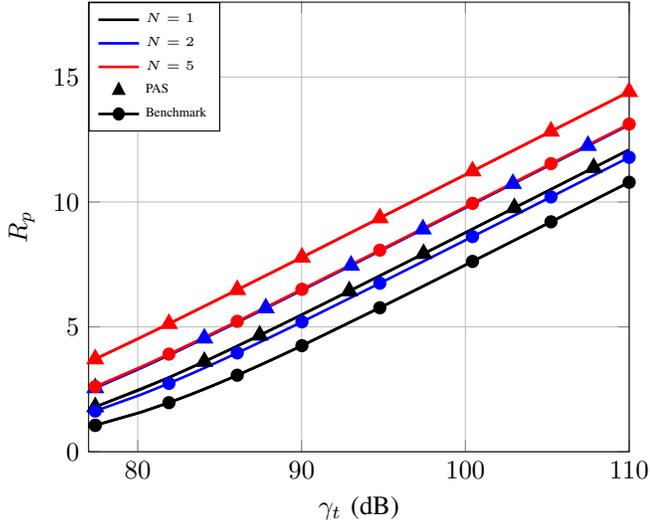

Finally, Fig. \ref{Fig6} shows the average rate as a function of $\gamma_t$ for a PAS with $N$ pinching antennas and a conventional system with $N$ colocated antennas. The results demonstrate that the PAS consistently outperforms the conventional system in all examined scenarios, highlighting its ability to leverage the flexibility of dynamically repositioning antennas along the waveguide to enhance communication performance. Notably, the PAS with only two antennas achieves the same rate performance as the conventional system with five antennas, underscoring the significant advantage provided by the unique degrees of freedom offered by pinching antennas. This ability allows them to compensate for path loss more effectively than conventional fixed antennas, making them highly efficient for high-rate communications.

\section{Conclusion}

In this work, we provided a comprehensive analytical framework for evaluating the performance of PASs by deriving closed-form expressions for critical metrics, including outage probability and average rate, under both ideal and realistic waveguide conditions. In particular, the provided expressions can be effectively used to characterize the performance of PASs in both single and multiple pinching antenna scenarios, providing valuable insights into their scalability and deployment strategies. In addition, our numerical results revealed that waveguide losses play a significant role in shaping system performance, especially as the waveguide length increases, underscoring their importance to be included in future work. Nevertheless, the results also showed that placing the pinching antenna at the location that minimizes the Euclidean distance between the transmitter and receiver remains a robust strategy, especially under low to moderate waveguide losses, ensuring effective communication reliability. Finally, our analysis verified the consistent superiority of PASs over conventional systems in terms of both reliability and data rate performance, further highlighting their transformative potential to overcome the limitations of conventional systems. To this end, our work lays the foundation for future investigations into their multi-user capabilities, optimal deployment strategies, and integration into PWEs.

\appendices
\section{Calculation of $I_i$ and $I_j$ integrals}\label{App:A}
Below we provide the calculations for both $I_i$ and $I_j$ integrals:
\subsubsection{Integral $I_i$}
Taking into account $I_1$, the integral $I_i$ can be written as
\begin{equation}
    I_i = \frac{1}{D_x D_y}\int_{w_1}^{w_2} D_y - 2\sqrt{Ce^{-\alpha x_m}-h^2}\, dx_m,
\end{equation}
which can be rewritten as
\begin{equation}\label{18}
    I_i = \frac{1}{D_x}\int_{w_1}^{w_2}1 \,dx_m- \frac{2}{D_x D_y}\int_{z_1}^{z_2}\sqrt{Ce^{-\alpha x_m}-h^2}\,dx_m.
\end{equation}
Setting $u=\sqrt{Ce^{-\alpha x_m}-h^2}$, \eqref{18} can be rewritten as
\begin{equation}\label{19}
    I_i=\frac{w_2-w_1}{D_x} + \frac{4}{\alpha D_x D_y} \int_{\sqrt{Ce^{-\alpha w_1}-h^2}}^{\sqrt{Ce^{-\alpha w_2}-h^2}} \frac{u^2}{ u^2+h^2}\, du,
\end{equation}
which can be further manipulated and written as
\begin{equation}\label{20}
     I_i=\frac{w_2-w_1}{D_x} + \frac{4}{\alpha D_x D_y} \int_{\sqrt{Ce^{-\alpha w_1}-h^2}}^{\sqrt{Ce^{-\alpha w_2}-h^2}} 1- \frac{h^2}{ u^2+h^2}\, du .
\end{equation}
Finally, utilizing the equation $h {\mathrm{tan}}^{-1}\left(\frac{x}{h}\right)=\int \frac{h^2}{x^2+ h^2}\, dx,$ \eqref{20} is written equivalently as in \eqref{I_i}, which concludes the calculation of $I_i$.

\subsubsection{Integral $I_j$}
Initially, the integral $I_j$ can be written as
\begin{equation}
    I_{j}=\frac{1}{D_x D_y}\int_{\frac{-1}{\alpha} \mathrm{ln}\left( \frac{h^2}{C}\right)}^{D_x} D_y \, dx_m,
\end{equation}
and thus, after some algebraic manipulations, we derive \eqref{I_j}, which concludes the calculation of $I_j$.

\section{Calculation of $I_A$ and $I_B$ integrals}\label{App:B}
Below we provide the calculations for both $I_A$ and $I_B$ integrals:
\subsubsection{Integral $I_A$}
First, by setting $w=\frac{A e^{-\alpha x_m}}{h^2 + \frac{D^2_y}{4}}$, we can rewrite \eqref{IA0} as
\begin{equation}\label{IA1}
    I_A= \frac{-1}{\alpha} \int_{w_1}^{w_2} \frac{\ln \left( 1+w\right)}{w} \,dw,
\end{equation}
where $w_1=\frac{A}{h^2 + \frac{D^2_y}{4}}$ and $w_2=\frac{A e^{-\alpha D_x}}{h^2 + \frac{D^2_y}{4}}$, respectively. Moreover, by setting $u=-w$, \eqref{IA1} can be rewritten as
\begin{equation}\label{IA3}
    I_A= -\frac{1}{\alpha} \int_{-w_1}^{-w_2} \frac{\ln \left( 1-u\right)}{u} \,du .
\end{equation}
Finally, by rewriting \eqref{IA3} as
\begin{equation}
        I_A= \frac{-1}{\alpha}\left(- \int_{0}^{-w_1} \frac{\ln \left( 1-u\right)}{u} \,du +\int_{0}^{-w_2} \frac{\ln \left( 1-u\right)}{u} \,du\right) ,
\end{equation}
and utilizing the equation $\mathrm{L_{i,2}}\left( x\right)= -\int_0^x \frac{\ln\left(1-u\right)}{u}du$, \eqref{IAt} can be derived, which concludes the calculation of $I_A$.
\subsubsection{Integral $I_B$}
Setting $\omega=\sqrt{Ae^{-\alpha x_m}+h^2}$, \eqref{IB0} can be written as
\begin{equation}\label{IB1}
    I_B= -\frac{2}{\alpha} \int_{\omega_1}^{\omega_2} \left(1+\frac{h^2}{z^2-h^2} \right) \tan^{-1}\!\left(\frac{D_y}{2\omega}\right) \, d\omega,
\end{equation}
where $\omega_1=\sqrt{A+h^2}$ and $\omega_2=\sqrt{Ae^{-\alpha D_x}+h^2}$, respectively. Therefore, \eqref{IB1} can be rewritten as
\begin{equation}\label{IB3}
\small
     I_B= \frac{-2}{\alpha} \left( \int_{\omega_1}^{\omega_2} \tan^{-1}\!\left(\frac{D_y}{2\omega}\right) \, d\omega + \int_{\omega_1}^{\omega_2} \frac{h^2   \tan^{-1}\!\left(\frac{D_y}{2\omega}\right)}{z^2-h^2} \, d\omega\right).
\end{equation}
Moreover, setting $v=\frac{D_y}{2 \omega}$, \eqref{IB3} can be written as
\begin{equation}\label{IB4}
\small
    \begin{split}
        &I_B= \frac{-2}{\alpha} \Bigg( \frac{-D_y}{2}\int_{v_1}^{v_2} \frac{ \tan^{-1}\!\left(v\right)}{v^2} \, dv + \int_{\omega_1}^{\omega_2} \frac{h^2   \tan^{-1}\!\left(\frac{D_y}{2\omega}\right)}{z^2-h^2} \, d\omega\Bigg),
    \end{split}
\end{equation}
where $v_1=\frac{D_y}{2\omega_1}$ and $v_2=\frac{D_y}{2\omega_2}$. Furthermore, performing integration by parts to both integrals, we can further rewrite \eqref{IB4} as
\begin{equation}\label{IB5}
\small
    \begin{split}
        &I_B= \frac{-2}{\alpha} \left(\! \frac{-D_y}{2} \left( \int_{v_1}^{v_2} \frac{1}{v\left( 1+v^2\right)}dv + \left[-\frac{\tan^{-1}\!\left(v\right)}{v}\right]_{v_1}^{v_2} \right) \right.\\
        &\!+\left. 
     \int_{\omega_1}^{\omega_2} \frac{h D_y \ln\left( \frac{\omega-h}{\omega+h}\right)}{4\left(\omega^2+\frac{D^2_y}{4}\right)}d\omega + \left[\frac{h\tan^{-1}\!\left(\frac{D_y}{2\omega}\right)}{2} \ln\left( \frac{\omega-h}{\omega+h}\right)\right]_{\omega_1}^{\omega_2} \right).
    \end{split}
\end{equation}
To further manipulate \eqref{IB5}, we can perform partial fraction decomposition to the two integrals. Therefore, after some algebraic manipulations, \eqref{IB5} is converted into
\begin{equation}\label{IB6}
\small
    \begin{split}
        &I_B = \frac{-2}{\alpha} \Bigg( 
        \frac{-D_y}{2} \left( \left[-\frac{\tan^{-1}\!\left(v\right)}{v}\right]_{v_1}^{v_2} +\int_{v_1}^{v_2} \frac{1}{v} - \frac{v}{1+v^2} \, dv 
         \right) \\
        &\!\!+ \left[\frac{h\tan^{-1}\!\left(\frac{D_y}{2\omega}\right)}{2} 
        \ln\left( \frac{\omega-h}{\omega+h}\right)\right]_{\omega_1}^{\omega_2} \!\!+ \frac{h}{4j} \Bigg( 
        \int_{\omega_1}^{\omega_2} \frac{\ln(\omega-h)}{\omega-j\frac{D_y}{2}} d\omega \\&
        \!\!- \!\!\int_{\omega_1}^{\omega_2} \frac{\ln(\omega+h)}{\omega-j\frac{D_y}{2}} d\omega -\!\! \int_{\omega_1}^{\omega_2} \frac{\ln(\omega-h)}{\omega+j\frac{D_y}{2}} d\omega + \!\!\int_{\omega_1}^{\omega_2} \frac{\ln(\omega+h)}{\omega+j\frac{D_y}{2}} d\omega \Bigg) \!\Bigg).
    \end{split}
\end{equation}
Thus, utilizing \cite[(2.01/2)]{GradshteynRyzhik2014} and \cite[(2.02/8)]{GradshteynRyzhik2014}, after some algebraic manipulations, we obtain
\begin{equation}\label{IB7}
\small
\begin{split}
    &I_B= \frac{-2}{\alpha} \Bigg[ \frac{D_y\ln\left(\frac{D^2_y}{4}+x^2\right)   }{4}+\frac{h\tan^{-1}\!\left(\frac{D_y}{2x} \right) }{2} \ln \left(\frac{x-h}{x+h} \right) \\&\!+x\tan^{-1}\!\left(\frac{D_y}{2x}\right)\Bigg]_{\omega_1}^{\omega_2} \!\!+ \frac{h}{4i} \Bigg( 
        \int_{\omega_1}^{\omega_2} \frac{\ln(\omega-h)}{\omega-j\frac{D_y}{2}} d\omega \\&
        \!\!-\!\! \int_{\omega_1}^{\omega_2} \frac{\ln(\omega-h)}{\omega+j\frac{D_y}{2}} d\omega - \!\!\int_{\omega_1}^{\omega_2} \frac{\ln(\omega+h)}{\omega-j\frac{D_y}{2}} d\omega + \!\!\int_{\omega_1}^{\omega_2} \frac{\ln(\omega+h)}{\omega+j\frac{D_y}{2}} d\omega \Bigg) ,
\end{split}
\end{equation}
As can be observed in \eqref{IB7}, to derive \eqref{IBt}, we need to prove that $I^{-}_C=z(x,h)$ and $I^{+}_C=z(x,-h)$, where $I^{-}_C$ and $I^{+}_C$ are equal to
\begin{equation}\label{IC0}
\small
    I^{-}_C=\frac{1}{j}\left(\int_{\omega_1}^{\omega_2} \frac{\ln(\omega-h)}{\omega-j\frac{D_y}{2}} d\omega - \int_{\omega_1}^{\omega_2} \frac{\ln(\omega-h)}{\omega+j\frac{D_y}{2}} d\omega\right),
\end{equation}
and
\begin{equation}
    I^{+}_C=\!\!\int_{\omega_1}^{\omega_2} \frac{\ln(\omega+h)}{\omega-j\frac{D_y}{2}} d\omega + \!\!\int_{\omega_1}^{\omega_2} \frac{\ln(\omega+h)}{\omega+j\frac{D_y}{2}} d\omega.
\end{equation}
Therefore, setting $t=\omega-h$, \eqref{IC0} can be written as
\begin{equation}\label{IC1}
\small
    I^{-}_C=\frac{1}{j}\left(\int_{\omega_1-h}^{\omega_2-h} \frac{\ln(t)}{t+h-j\frac{D_y}{2}} dt - \int_{\omega_1-h}^{\omega_2-h} \frac{\ln(t)}{t+h+j\frac{D_y}{2}} dt\right).
\end{equation}
Moreover, utilizing \cite[(2.727/2)]{GradshteynRyzhik2014} and \cite[(2.728/2)]{GradshteynRyzhik2014}, \eqref{IC1} can be rewritten as
\begin{equation}\label{44}
\small
\begin{split}
I^{-}_C &= \frac{1}{j} \Bigg[ 
    \ln(t)\ln\left(\frac{t+h-j\frac{D_y}{2}}{t+h+j\frac{D_y}{2}}\right) 
    - \ln(t)\ln\left(\frac{h-j\frac{D_y}{2}}{h+j\frac{D_y}{2}}\right) \\
&+ \mathrm{L}_{i,2}\left(\frac{-t}{h-j\frac{D_y}{2}}\right) 
     - \mathrm{L}_{i,2}\left(\frac{-t}{h+j\frac{D_y}{2}}\right)
    \Bigg]_{\omega_1-h}^{\omega_2-h}.
\end{split}
\end{equation}
In addition, utilizing $\omega=t+h$ and the equation $\ln\left(\frac{x-ja}{x+ja}\right)=j\left(2 \tan^{-1}\!\left(\frac{x}{a}\right) -\pi\right)$, after some algebraic manipulations, we can write \eqref{44} as follows
\begin{equation}
\small
\begin{aligned}
    I^{-}_C &= \frac{1}{j} \Bigg[ 2j\ln\left(\omega-h\right) \left(\tan^{-1}\!\left(\frac{2\omega}{D_y}\right) - \tan^{-1}\!\left(\frac{2h}{D_y}\right)\right)  \\
    &+ \mathrm{L}_{i,2}\left(\frac{h-\omega}{h-j\frac{D_y}{2}}\right) 
     - \mathrm{L}_{i,2}\left(\frac{h-\omega}{h+j\frac{D_y}{2}}\right)
    \Bigg]_{\omega_1}^{\omega_2}.
\end{aligned}
\end{equation}
Finally, considering that the arguments of the dilogarithm functions are complex numbers and conjugate $\forall \omega \in \left[\omega_2,\omega_1 \right]$, it holds that $\mathrm{L}_{i,2}\left(\frac{h-\omega}{h-j\frac{D_y}{2}}\right) -\mathrm{L}_{i,2}\left(\frac{h-\omega}{h+j\frac{D_y}{2}}\right)=2j \mathrm{Im}\left\{\mathrm{L}_{i,2}\left(\frac{h-\omega}{h-j\frac{D_y}{2}}\right)\right\}$, thus proving that $I^{-}_C=z(x,h)$. Therefore, by similarly proving that $I^{+}_C=z(x,-h)$,  after some algebraic manipulations, we obtain \eqref{IB0}, which concludes the calculation of $I_B$.


\bibliographystyle{IEEEtran}
\bibliography{Bibliography}

\begin{thebibliography}{10}
\providecommand{\url}[1]{#1}
\csname url@samestyle\endcsname
\providecommand{\newblock}{\relax}
\providecommand{\bibinfo}[2]{#2}
\providecommand{\BIBentrySTDinterwordspacing}{\spaceskip=0pt\relax}
\providecommand{\BIBentryALTinterwordstretchfactor}{4}
\providecommand{\BIBentryALTinterwordspacing}{\spaceskip=\fontdimen2\font plus
\BIBentryALTinterwordstretchfactor\fontdimen3\font minus \fontdimen4\font\relax}
\providecommand{\BIBforeignlanguage}[2]{{%
\expandafter\ifx\csname l@#1\endcsname\relax
\typeout{** WARNING: IEEEtran.bst: No hyphenation pattern has been}%
\typeout{** loaded for the language `#1'. Using the pattern for}%
\typeout{** the default language instead.}%
\else
\language=\csname l@#1\endcsname
\fi
#2}}
\providecommand{\BIBdecl}{\relax}
\BIBdecl

\bibitem{Holographic}
I.~F. Akyildiz and H.~Guo, ``Holographic-type communication: {A} new challenge for the next decade,'' \emph{ITU J. Future Evolving Technol.}, vol.~3, no.~2, pp. 421--442, Sep. 2022.

\bibitem{healthcare}
\BIBentryALTinterwordspacing
A.~Ahad, Z.~Jiangbina, M.~Tahir, I.~Shayea, M.~A. Sheikh, and F.~Rasheed, ``{6G} and intelligent healthcare: {Taxonomy}, technologies, open issues and future research directions,'' \emph{Internet Things}, vol.~25, p. 101068, 2024. [Online]. Available: \url{https://www.sciencedirect.com/science/article/pii/S2542660524000106}
\BIBentrySTDinterwordspacing

\bibitem{6GMetaverse}
H.~Yu, M.~Shokrnezhad, T.~Taleb, R.~Li, and J.~Song, ``Toward {6G}-based metaverse: {Supporting} highly-dynamic deterministic multi-user extended reality services,'' \emph{IEEE Netw.}, vol.~37, no.~4, pp. 30--38, 2023.

\bibitem{6GNetwork}
H.~Jiang, M.~Mukherjee, J.~Zhou, and J.~Lloret, ``Channel modeling and characteristics for {6G} wireless communications,'' \emph{IEEE Netw.}, vol.~35, no.~1, pp. 296--303, 2021.

\bibitem{PWELiaskos}
C.~Liaskos, A.~Tsioliaridou, S.~Nie, A.~Pitsillides, S.~Ioannidis, and I.~F. Akyildiz, ``On the network-layer modeling and configuration of programmable wireless environments,'' \emph{IEEE/ACM Trans. Netw.}, vol.~27, no.~4, pp. 1696--1713, 2019.

\bibitem{SRE2020}
M.~Di~Renzo, A.~Zappone, M.~Debbah, M.-S. Alouini, C.~Yuen, J.~de~Rosny, and S.~Tretyakov, ``Smart radio environments empowered by reconfigurable intelligent surfaces: {How} it works, state of research, and the road ahead,'' \emph{IEEE J. Sel. Areas Commun.}, vol.~38, no.~11, pp. 2450--2525, 2020.

\bibitem{liaskosmagazine}
C.~Liaskos, S.~Nie, A.~Tsioliaridou, A.~Pitsillides, S.~Ioannidis, and I.~Akyildiz, ``A new wireless communication paradigm through software-controlled metasurfaces,'' \emph{IEEE Commun. Mag.}, vol.~56, no.~9, pp. 162--169, 2018.

\bibitem{EEAlexand}
C.~Huang, A.~Zappone, G.~C. Alexandropoulos, M.~Debbah, and C.~Yuen, ``Reconfigurable intelligent surfaces for energy efficiency in wireless communication,'' \emph{IEEE Trans. Wireless Commun.}, vol.~18, no.~8, pp. 4157--4170, 2019.

\bibitem{TegosTVT}
S.~A. Tegos, D.~Tyrovolas, P.~D. Diamantoulakis, C.~K. Liaskos, and G.~K. Karagiannidis, ``On the distribution of the sum of double-{Nakagami-$m$} random vectors and application in randomly reconfigurable surfaces,'' \emph{IEEE Trans. Veh. Technol.}, vol.~71, no.~7, pp. 7297--7307, 2022.

\bibitem{zeris}
D.~Tyrovolas, S.~A. Tegos, V.~K. Papanikolaou, Y.~Xiao, P.-V. Mekikis, P.~D. Diamantoulakis, S.~Ioannidis, C.~K. Liaskos, and G.~K. Karagiannidis, ``Zero-energy reconfigurable intelligent surfaces {(zeRIS)},'' \emph{IEEE Trans. Wireless Commun.}, vol.~23, no.~7, pp. 7013--7026, 2024.

\bibitem{Active}
Z.~Zhang, L.~Dai, X.~Chen, C.~Liu, F.~Yang, R.~Schober, and H.~V. Poor, ``Active {RIS} vs. passive {RIS}: {Which} will prevail in {6G}?'' \emph{IEEE Trans. Commun.}, vol.~71, no.~3, pp. 1707--1725, 2023.

\bibitem{STARRIS}
J.~Xu, Y.~Liu, X.~Mu, and O.~A. Dobre, ``{STAR-RISs}: {Simultaneous} transmitting and reflecting reconfigurable intelligent surfaces,'' \emph{IEEE Commun. Lett.}, vol.~25, no.~9, pp. 3134--3138, 2021.

\bibitem{BDRIS}
H.~Li, S.~Shen, M.~Nerini, and B.~Clerckx, ``Reconfigurable intelligent surfaces 2.0: {Beyond} diagonal phase shift matrices,'' \emph{IEEE Commun. Mag.}, vol.~62, no.~3, pp. 102--108, 2024.

\bibitem{LERIS}
D.~Tyrovolas, D.~Bozanis, S.~A. Tegos, V.~K. Papanikolaou, P.~D. Diamantoulakis, C.~K. Liaskos, R.~Schober, and G.~K. Karagiannidis, ``Empowering programmable wireless environments with optical anchor-based positioning,'' \emph{IEEE Netw.}, vol.~39, no.~1, pp. 14--20, 2025.

\bibitem{Fluid}
K.-K. Wong, A.~Shojaeifard, K.-F. Tong, and Y.~Zhang, ``Fluid antenna systems,'' \emph{IEEE Trans. Wireless Commun.}, vol.~20, no.~3, pp. 1950--1962, 2021.

\bibitem{Movable}
L.~Zhu, W.~Ma, and R.~Zhang, ``Movable antennas for wireless communication: {Opportunities} and challenges,'' \emph{IEEE Commun. Mag.}, vol.~62, no.~6, pp. 114--120, 2024.

\bibitem{XRRF}
C.~Liaskos \emph{et~al.}, ``{XR-RF} imaging enabled by software-defined metasurfaces and machine learning: {Foundational} vision, technologies and challenges,'' \emph{IEEE Access}, vol.~10, pp. 119\,841--119\,862, 2022.

\bibitem{ISACRIS}
F.~Jiang, A.~Abrardo, K.~Keykhosravi, H.~Wymeersch, D.~Dardari, and M.~Di~Renzo, ``Two-timescale transmission design and {RIS} optimization for integrated localization and communications,'' \emph{IEEE Trans. Wireless Commun.}, vol.~22, no.~12, pp. 8587--8602, 2023.

\bibitem{book}
C.~Yeh and F.~I. Shimabukuro, \emph{The Essence of Dielectric Waveguides}.\hskip 1em plus 0.5em minus 0.4em\relax Springer, 2008.

\bibitem{DOCOMO}
A.~Fukuda, H.~Yamamoto, H.~Okazaki, Y.~Suzuki, and K.~Kawai, ``Pinching antenna-using a dielectric waveguide as an antenna,'' \emph{NTT DOCOMO Tech. J.}, vol.~23, no.~3, pp. 5--12, Jan. 2022.

\bibitem{pinchingMAG}
\BIBentryALTinterwordspacing
Z.~Yang, N.~Wang, Y.~Sun, Z.~Ding, R.~Schober, G.~K. Karagiannidis, V.~W.~S. Wong, and O.~A. Dobre, ``Pinching antennas: {Principles}, applications and challenges,'' 2025. [Online]. Available: \url{https://arxiv.org/abs/2501.10753}
\BIBentrySTDinterwordspacing

\bibitem{Ding2024TCOM}
\BIBentryALTinterwordspacing
Z.~Ding, R.~Schober, and H.~V. Poor, ``Flexible-antenna systems: {A} pinching-antenna perspective,'' 2024. [Online]. Available: \url{https://arxiv.org/abs/2412.02376}
\BIBentrySTDinterwordspacing

\bibitem{TegosPinching}
\BIBentryALTinterwordspacing
S.~A. Tegos, P.~D. Diamantoulakis, Z.~Ding, and G.~K. Karagiannidis, ``Minimum data rate maximization for uplink pinching-antenna systems,'' 2024. [Online]. Available: \url{https://arxiv.org/abs/2412.13892}
\BIBentrySTDinterwordspacing

\bibitem{NOMAPinch}
\BIBentryALTinterwordspacing
K.~Wang, Z.~Ding, and R.~Schober, ``Antenna activation for {NOMA} assisted pinching-antenna systems,'' 2024. [Online]. Available: \url{https://arxiv.org/abs/2412.13969}
\BIBentrySTDinterwordspacing

\bibitem{PASS}
\BIBentryALTinterwordspacing
C.~Ouyang, Z.~Wang, Y.~Liu, and Z.~Ding, ``Array gain for pinching-antenna systems {(PASS)},'' 2025. [Online]. Available: \url{https://arxiv.org/abs/2501.05657}
\BIBentrySTDinterwordspacing

\bibitem{GradshteynRyzhik2014}
I.~S. Gradshteyn and I.~M. Ryzhik, \emph{Table of Integrals, Series, and Products}.\hskip 1em plus 0.5em minus 0.4em\relax New York, NY, USA: Academic Press, 2014.

\end{thebibliography}

\end{document}